\documentclass[aip,jmp,amsmath,amssymb,preprint,
author-numerical]{revtex4-1}
\usepackage{dcolumn}
\usepackage{bm}
\usepackage[mathlines]{lineno}
\usepackage{epsfig}
\usepackage{epstopdf}
\usepackage{bm}
\usepackage{float}
\usepackage{subfig}
\usepackage[bookmarks=false]{hyperref}

\begin{document}
	
	\title{
		Nonlocal Orbital-Free kinetic pressure tensors for the Fermi gas}
		
	
	\author{D. I. Palade}
	\email{dragos.palade@inflpr.ro}
	\affiliation{ National Institute of Laser, Plasma and Radiation Physics,
		PO Box MG 36, RO-077125 M\u{a}gurele, Bucharest, Romania }
	\affiliation{Faculty of Physics, University of Bucharest, Romania
	}%
	
	\date{\today}
	
	\begin{abstract}
     

A novel nonlocal density functional for the kinetic pressure tensor of a Fermi gas is derived. The functional is designed to reconcile the Quantum Hydrodynamic Model with the microscopic approaches, both for homogeneous equilibrium and dynamical regime. The derivation opens new ways to improve and implement further time-nonlocal functionals. The present proposal is systematically tested in and beyond the linear regime for the Fermi gas, as well as for some small sodium clusters, proving that it is quantitative superior to other existing functionals.
		
	\end{abstract}
	
	\maketitle

\section{Introduction}	
\label{Sec. I}


Systems containing (partially) degenerate fermionic species (warm-dense matter \cite{Fletcher2015}, nano-particles\cite{Scholl2012}, metallic clusters \cite{CALVAYRAC2000493}, semiconductors, thin metal films\cite{PhysRevB.78.155412}, dense astrophysical objects\cite{doi:10.1002/ctpp.201300001}, etc.) have drawn great interest in the past decades, especially due to recent experimental and technological progress. In particular, nano-systems exhibiting quantum plasmonic behavior became one of the paradigms for future nano-electronic devices \cite{Tame2013,Schuller2010} due to their ability to enhance and localize electromagnetic radiation bellow the diffraction limit\cite{Gramotnev2010}.

In general, such systems contain an ionic and one or more fermionic (electrons in metals, electrons and holes in semiconductors) species. Due to their large inertia and localized spatial distribution, ions can be safely considered as being purely classical objects in most scenarios. The fermions, especially at low temperatures and high densities, display strong quantum features following closely the Fermi-Dirac statistic. Naturally, quantum theoretical methods are required to describe the physics behind quantum Fermi systems, both at equilibrium and during their dynamics.

In practice, kinetic (quantum Wigner\cite{0295-5075-88-3-35001,HAAS2010481}) and microscopic (Hartree-Fock like\cite{9780124865525}, TD-DFT\cite{PhysRevLett.52.997}) theories offer high precision to numerical complexity ratio. Unfortunately, the numerical complexity of microscopic approaches scales with the number of particles, while the kinetic approaches involve 6+1 dimensional partial differential equations. For example, investigating within DFT large metal clusters ( $\sim 10^{1-2}nm$ and  $N\sim 10^{3-4}$ particles) in full 3D geometry remains a prohibitive numerical task even with the new generation of CPU processors. 

In the given context, a simpler model has gained recognition in the past decades: the Quantum Hydrodynamic Model (QHD)\cite{doi:10.1137/S0036139992240425,PhysRevB.78.155412}. Although in literature it appears under different names: (Time-Dependent-)Thomas-Fermi \cite{PhysRevLett.80.5520,doi:10.1002/ctpp.201500024,0953-4075-48-18-185102}, Quantum Hydrodynamic Theory\cite{doi:10.1063/1.5003910} or Quantum Fluid Theory\cite{PhysRevB.64.075316}, it mainly consists of two conservation laws: a continuity equation for the total density of particles $n(\mathbf{r},t)$ and a momentum equation for the total density of current $\mathbf{j}(\mathbf{r},t)$ (alternatively, the average velocity field  $\mathbf{u}(\mathbf{r},t)= \mathbf{j}(\mathbf{r},t)/n(\mathbf{r},t)$). The model has been applied with a fare amount of success to a variety of systems: nuclei \cite{MYERS1996141,CHU1977407}, atomic and molecular systems \cite{PhysRevA.46.5130}, metallic clusters \cite{PhysRevLett.80.5520,0953-4075-48-18-185102}, quantum plasmas \cite{RevModPhys.83.885,doi:10.1063/1.4958324}, etc.

Similar to classical hydrodynamics, the QHD lacks a closure relation between the kinetic pressure tensor $\hat{\Pi}(\mathbf{r},t)$ and the other lower moments of the distribution function: density and current. Within the Density Functional Theory (DFT), in particular through the Runge-Ross theorems\cite{PhysRevLett.52.997}, it can be shown that this tensor is an unique (exact and unknown) density functional $\hat{\Pi}(\mathbf{r},t)\equiv \hat{\Pi}[n(\mathbf{r},t)]$. 

The branch of DFT concerned with this universal functional (or with better approximations of it) is known as Orbital-Free-DFT (OF-DFT)\cite{KARASIEV20122519} and it dates back to the foundations of quantum physics. Alternatively, OF-DFT deals with functionals for the kinetic energy which is equal with the trace of the kinetic pressure tensor. Despite being a long-standing problem, most applications of the QHD still use a functional developed almost a century ago, the Thomas-Fermi(-Bohm) approximation (also known as Thomas-Fermi-von-Weizsacker\cite{Benguria1981}):
\begin{equation}
\label{Eq_1}
\hat{\Pi}[n]=P_{TF}[n]\hat{1}-\lambda\frac{\hbar^2}{4m^2}n\nabla\otimes\nabla\ln n.\end{equation}
with $P_{TF}[n_0]=2E_F[n_0]n_0/5m$ the Thomas-Fermi pressure, $\hat{1}$ the identity tensor, while the second term is a reformulation of the Bohm potential\cite{PhysRev.85.166} which we shall refer to as macroscopic Bohm pressure. Historically, the $\lambda$ constant spanned the $[0,1]$ interval. It has been emphasized\cite{Kirzhnits1957,doi:10.1002/ctpp.201500024} and generally accepted that $\lambda=1$ for bosonic and $\lambda=1/9$ for fermionic systems (in 3D at $T=0K$). A detailed discussion on this matter will be presented throughout this work.

The TF-Bohm approximation belongs to a more general scheme named gradient-expansion\cite{PhysRevB.88.161108}. Being derived from the microscopic equilibrium of the homogeneous electron gas (HEG), all these schemes are valid only for the nearly-free stationary Fermi gas. Trying to recreate simple dynamic phenomena such as the propagation of an electrostatic wave through a HEG results in the impossibility of TF-Bohm to reproduce (within the QHD) even the dispersion relation in the long wavelength limit\cite{PhysRevB.91.115416}. The previous example hides a very serious pathology: the kinetic/microscopic dynamic behavior of fermionic systems is not captured by the functional \ref{Eq_1}. From this point of view, the limitations of QHD are surprisingly rare brought into attention\cite{PhysRevB.93.205405,PhysRevB.91.115416,doi:10.1063/1.4907167,doi:10.1063/1.5003910}. The general recipe is to use the approximation [\ref{Eq_1}] with $\lambda = 1/9$ for equilibrium configurations while, in the linear regime, the functional is modified to:
$$\left(\frac{\delta \hat{\Pi}}{\delta n}\right)=\frac{9}{5}\left(\frac{\delta P_{TF}[n]}{\delta n}\right)+\lambda\left(\frac{\delta \hat{\Pi}_B[n]}{\delta n}\right).$$
with $\lambda = 1$. This scheme is designed to work only in the limit of high frequency and short wavelengths, its mathematical inconsistency being undeniable evidence that TF-Bohm (and its extensions\cite{PhysRevB.88.161108}) are invalid during dynamics. Given these limitations, there should be serious doubts regarding the results of TF-Bohm applications in fully non-linear regimes dominated by wave-mixing on a wide spectrum of frequencies and wave-numbers\cite{PhysRevLett.96.245001,PhysRevLett.99.125002,RevModPhys.83.885}.

One of the first solutions\cite{PhysRevB.60.16350,PhysRevB.81.045206} to the inaccuracy of the TF-Bohm functional was designed for the equilibrium configurations of metallic and semiconductor systems. The main proposal was that a density functional should have a non-local character in space in order to reproduce the static linear response function. Recently\cite{doi:10.1063/1.5003910,doi:10.1002/ctpp.201700113} the idea has been extended, stressing that a density functional for the free energy of an electron gas should reproduce within QHD the dynamic linear response function (Lindhard, RPA polarization function). 

The functionals derived in\cite{doi:10.1063/1.5003910,doi:10.1002/ctpp.201700113} are designed to reconcile \emph{linearized} QHD with the kinetics of an electron gas in asymptotic spectral regions, low ($\omega\ll \hbar k^2/m$) or high ($\omega\gg \hbar k^2/m$) frequency and include non-zero temperature effects. At $T=0K$ they read $\hat{\Pi}^0_{\alpha,\lambda}[n]=\alpha\hat{\Pi}_{TF}[n]+\lambda\hat{\Pi}_B[n]$ where:
$$(\alpha, \beta) = \left\{
\begin{array}{lll}
	(1,\frac{1}{9}) & \mbox{  for  } \omega\ll \hbar k^2/2m,\hspace{0.1cm}k\ll k_F,\\
	(\frac{3}{5},1) & \mbox{  for  } \omega\ll \hbar k^2/2m,\hspace{0.1cm}k\gg k_F,\\
    (\frac{9}{5},1) & \mbox{  for  } \omega\gg \hbar k^2/2m.
\end{array}
\right.
$$
Their most important feature is the analytical simplicity which passes as facility in numerical implementation. Their interpretation, non-local Bohm potential, is disproved in the present work. More important, the lack of an unified expression over all spectral regions makes these functionals useless for physical scenarios in which high and low frequency modes coexist during dynamics on a wide range of wavelengths\cite{PhysRevLett.96.245001,PhysRevLett.99.125002}. This is almost always the case for systems relevant to nano-plasmonics in non-linear regimes when high frequency electrostatic waves are present simultaneously with low frequency ion dynamics and, sometimes, short wavelength density oscillations. Finally, there are situations when the assumption that $\nabla\cdot\hat{\Pi}[n]=n\nabla (\delta T[n]/\delta n)$ where $T[n]$ is the kinetic energy functional is not true, therefore, invalidating a field theoretical description. For these reasons, the goal of the present work is to construct a time-nonlocal kinetic pressure density functional (KPDF) which can reproduce concurrently the equilibrium as well as the dynamical linear configurations of a Fermi gas.

The paper is organized as it follows: in Section\ref{Sec. II} the problem is posed in the contexts of TD-DFT and quantum Wigner equation. From microscopic analysis, the form of the KPDF is motivated and, imposing the associated constrains for equilibrium and linear response, it is explicitly derived. Through a reasonable approximation, the time-nonlocality of the functional is reformulated as a wave-like equation. In Section \ref{Sec. III} the validity of the proposed approximation is investigated from a spectral perspective. Extensive comparisons between the KPDF and the  microscopic results are performed for the dynamics of a Fermi gas. Finally, realistic small sodium clusters are simulated to establish the qualitative advantages of the KPDF over existing approximations. 

\section{Theory}
\label{Sec. II}

\subsection{Framework}
\label{Sec. IIa}
Let us consider a $N$-body fermionic system. For simplicity, relativistic, temperature, spin or magnetic effects are neglected. Although not involved in the derivation of the functional, a two-body interaction (e.g. Coulomb) is allowed in addition to an external potential $v_{ext}$. The microscopic description offered by TD-DFT \cite{PhysRevLett.52.997} assigns to each particle a pseudo-orbital $\psi_k(\mathbf{r},t), \forall k=1,N$ which obeys the Kohn-Sham [KS] equations:
\begin{eqnarray}\label{Eq_2}
i\hbar\partial_t|\psi_k\rangle=\hat{H}|\psi_k\rangle\\
\hat{H}=\frac{\hat{p}^2}{2m}+\hat{v}_{eff}\nonumber
\end{eqnarray}
where $\hat{H}$ is the single-particle Hamiltonian operator, $m$ is the fermionic mass, while $v_{eff}$ is the effective potential which, aside from the external potential, includes a mean-field interaction and an exchange-correlation term \cite{PhysRev.136.B864}  $v_{eff}=v_{ext}+v_{mf}+v_{xc}$. The ground state is subject to an eigenvalue problem: $\varepsilon_k|\psi_k\rangle=\hat{H}_0|\psi_k^0\rangle$.

An alternative description can be rendered either starting from the $N$-body quantum Liouville equation all the way through a BBGKY hierarchy\cite{9783319241210}, or by defining the single-particle density operator from the KS orbitals $\hat{\rho}=\sum_kp_k|\psi_k\rangle\langle\psi_k|$. Both ways deliver the so called quantum Wigner equation\cite{0295-5075-88-3-35001,HAAS2010481} for $\hat{\rho}$:
\begin{equation}\label{Eq_3}
i\hbar\partial_t\hat{\rho}=[\hat{H},\hat{\rho}].\end{equation}
Within the position representation, one can define hydrodynamic quantities: total density of particles,
$$n(\mathbf{r},t)=\sum_kp_k|\psi_k(\mathbf{r},t)|^2=\lim\limits_{\mathbf{r}'\to \mathbf{r}}\rho(\mathbf{r},\mathbf{r}',t),$$
total current density,
$$\mathbf{j}(\mathbf{r},t)=\sum_kp_k\mathbf{j}_k(\mathbf{r},t)=\frac{\hbar}{2mi}\lim\limits_{\mathbf{r}'\to \mathbf{r}}(\nabla_\mathbf{r}-\nabla_{\mathbf{r}'})\rho(\mathbf{r},\mathbf{r}',t)$$
and the kinetic pressure tensor $\hat{\Pi}(\mathbf{r},t)$:
\begin{align}\label{Eq_4}\nonumber
\hat{\Pi}_{tot}(\mathbf{r},t)=-\frac{\hbar^2}{4m^2}\sum_kp_kn_k\nabla\otimes\nabla\ln n_k+\sum_kp_k\frac{\mathbf{j}_k\otimes \mathbf{j}_k}{n_k}=\\
=-\frac{\hbar^2}{4m^2}\lim\limits_{\mathbf{r}'\to \mathbf{r}}(\nabla_\mathbf{r}-\nabla_\mathbf{r'})\otimes(\nabla_\mathbf{r}-\nabla_\mathbf{r'})\rho(\mathbf{r},\mathbf{r}',t)
\end{align}
with $n_k=|\psi_k|^2$ the single particle density and $\mathbf{j}_k=\hbar/2mi (\psi_k^*\nabla\psi_k-\psi_k\nabla\psi_k^*)$ the single particle current. Either starting from the microscopic KS Eqns. [\ref{Eq_2}] together with a Madelung representation of orbitals $\psi_k=n_k^{1/2}exp(iS_k/\hbar)$, or simply by using the Wigner Eq. [\ref{Eq_3}], one can derive \cite{PhysRevB.64.075316,PhysRevB.78.155412} the Quantum Hydrodynamic Model:
\begin{align}\label{Eq_5a}
&\partial_tn+\nabla j=0\\
\label{Eq_5b}
&\partial_t\mathbf{j}+\nabla(\frac{\mathbf{j}\otimes \mathbf{j}}{n})+\frac{n}{m}\nabla v_{eff}+\nabla\cdot\hat{\Pi}=0.
\end{align}

In the momentum equation $\hat{\Pi}\equiv\hat{\Pi}_{tot}-\mathbf{j}\otimes\mathbf{j}/n$ stands for the reduced kinetic pressure tensor which is exactly the topic of the present paper. The Runge-Ross theorems prescribe \cite{PhysRevLett.52.997} that the system [\ref{Eq_5a},\ref{Eq_5b}] is valid and $\hat{\Pi}\equiv\hat{\Pi}[n]$. For future purposes, let us decompose $\hat{\Pi}_{tot}$ in the \emph{microscopic Thomas-Fermi pressure} $\hat{\mathcal{P}}_{TF}$ and the \emph{microscopic Bohm} pressure $\hat{\mathcal{P}}_B$:
\begin{align}\label{Eq_6a}
&\hat{\mathcal{P}}_{TF}=\sum_kn_k\mathbf{u}_k\otimes\mathbf{u}_k\\
\label{Eq_6b}
&\hat{\mathcal{P}}_{B}=-\frac{\hbar^2}{4m^2}\sum_k n_k\nabla\otimes\nabla\ln n_k
\end{align}
where $\mathbf{u}_k=\mathbf{j}_k/n_k$ is the single particle velocity field. These definitions are consistent with the historical \cite{PhysRev.96.208} prescription of the Bohm potential. 

\subsection{Two Fermi systems}
\label{Sec. IIb}
Let us consider two similar systems $\mathcal{S}^1$ and $\mathcal{S}^2$, both depicting the $N$-body $d$ dimensional \emph{homogeneous Fermi gas} ($v_{eff}=0$) : $\mathcal{S}^1$ with periodic and $\mathcal{S}^2$ with null Dirichlet boundary conditions on a box of length $L$. $\mathcal{S}^1$ describes free particles (continuum fermions) while $\mathcal{S}^2$ describes fully confined fermions. The microscopic stationary orbitals read: 
\begin{eqnarray}\nonumber
&\psi_k^{\mathcal{S}^1}(\mathbf{r},t=0)=L^{-d/2}e^{-i\mathbf{k}\mathbf{r}}\\
&\psi_k^{\mathcal{S}^2}(\mathbf{r},t=0)=(L/2)^{-d/2}\prod_{j=1}^{d}\sin(k_jx_j)\nonumber
\end{eqnarray}
with $\mathbf{k}=2\pi/L\mathbf{n}$. By means of Madelung representation, in $\mathcal{S}^1$: $n_k(\mathbf{r},t)=L^{-d/2}$ and $S_k(\mathbf{r},t)=\mathbf{k}\cdot\mathbf{r}$
while in $\mathcal{S}^2$: $n_k(\mathbf{r},t)=(L/2)^{-d}\sin(k_jx_j)^2$ and $S_k(\mathbf{r},t)=0$. This can be interpreted as it follows: in $\mathcal{S}^1$ the densities $n_k$ have bosonic behavior (the same values $\forall k$) while the phases $S_k$ fermionic behavior (uniform distribution). The converse is true in $\mathcal{S}^2$.

All macroscopic quantities, both at equilibrium, as well as in linear regime (under the influence of a small effective potential $\delta v_{eff}(\mathbf{r},t)$) can be computed analytically. Table \ref{table0} shows, as complete as possible, these results ($P_{TF}=2E_F[n_0]n_0/5m$, the Fermi energy $E_F=\hbar^2k_F^2/2m$ and the Fermi wavevector $k_F=(3\pi^2n_0)^{1/3}$). 
\begin{center}\label{table0}
	\begin{tabular}{ |c|c|c| } 
		\hline
		& $\mathcal{S}^1$ & $\mathcal{S}^2$ \\ 
		\hline
		$n(\mathbf{r},t=0)$ & $n_0$ & $n_0$ \\ 
		\hline
		$\mathbf{j}(\mathbf{r},t=0)$ & 0 & 0 \\ 
		\hline
		$\hat{\Pi}(\mathbf{r},t=0)$ & $P_{TF}[n_0] \hat{1}$ & $P_{TF}[n_0] \hat{1}$ \\ 
		\hline
		$\hat{\mathcal{P}}_{TF}(\mathbf{r},t=0)$ & $P_{TF}[n_0] \hat{1}$ & $0$ \\ 
		\hline
		$\hat{\mathcal{P}}_B(\mathbf{r},t=0)$ & $0$ & $P_{TF}[n_0] \hat{1}$ \\ 
		\hline
		$\delta\hat{\Pi}(\mathbf{r},t)$ & $\delta\hat{\Pi}(\mathbf{r},t)$ & $\delta\hat{\Pi}(\mathbf{r},t)$ \\ 
		\hline
		$\delta\hat{\mathcal{P}}_B(\mathbf{r},t)$ & $-\frac{\hbar^2}{4m^2}\nabla\otimes\nabla\delta n$ & $-\frac{\hbar^2}{4m^2}\nabla\otimes\nabla\delta n+other$ \\ 
		\hline
		$\delta\hat{\mathcal{P}}_{TF}(\mathbf{r},t)$ & $\neq 0$ & $0$ \\ 
		\hline
	\end{tabular}
\captionof{table}{Macroscopic quantities and microscopic pressures for the systems $\mathcal{S}^{1,2}$ at homogeneous equilibrium and during linear dynamics.}
\end{center}

From the macroscopic perspective, both systems appear to be identical. Nonetheless, investigating the microscopic pressures at equilibrium, one can conclude that the macroscopic Thomas-Fermi pressure $\hat{\Pi}_{TF}[n]=P_{TF}[n]\hat{1}$ is universal whereas its nature is related to the fermionic nature of the system: in $\mathcal{S}^1$ the fermionic character is exhibited by phases (velocities) therefore, the microscopic TF reproduces the whole macroscopic TF. In $\mathcal{S}^2$ the converse is true, the macroscopic TF being reproduced by the microscopic Bohm. The presence of a macroscopic Bohm pressure $\hat{\Pi}_B=-\hbar^2/4m^2 n\nabla\otimes\nabla\ln n$ at equilibrium can be assumed, since the density is uniform and gives null contribution.
 
Within linear dynamics, the universality of the macroscopic Bohm term $\hat{\Pi}_B$ with a prefactor $\lambda=1$ spectrally independent is suggested by the presence of $-\hbar^2/4m^2\nabla\otimes\nabla\delta n$ for both systems. Moreover, the deviations of $\hat{\Pi}$ from a TF-Bohm functional cannot be assigned only to the microscopic Bohm term.

In general cases, expanding in Eqns.\ref{Eq_6a},\ref{Eq_6b} the velocities and the densities around their average $\mathbf{u}_k=\mathbf{u}_k'+\mathbf{u}$, $n_k=n/N+ n_k'$ yields:
\begin{align*}
&\hat{\mathcal{P}}_{TF}=\frac{\mathbf{j}\otimes\mathbf{j}}{n}+\hat{\mathcal{P}}_{TF}'\\
&\hat{\mathcal{P}}_{B}=\hat{\Pi}_B[n]+\hat{\mathcal{P}}_{B}'\\
\hat{\mathcal{P}}_{TF}'+&\hat{\mathcal{P}}_B'=\hat{\Pi}_{TF}[n]+\hat{\Pi}_{NL}[n].
\end{align*}

This decomposition underlines that the advection pressure $\mathbf{j}\otimes\mathbf{j}/n$ and $\hat{\Pi}_B$ are zero order average terms which should be universal. The macroscopic TF pressure is a consequence of the fermionic character encoded in both microscopic TF and Bohm terms. The residue up to the total real pressure tensor is denoted as $\Pi_{NL}=\hat{\Pi}-\hat{\Pi}_{TF}-\hat{\Pi}_B$ and will be later investigated. It must be emphasized that its origin is also the fermionic character of the system and the coupling between $\mathcal{P}_{TF}$ and $\mathcal{P}_B$. Previously \cite{doi:10.1002/ctpp.201700113,doi:10.1063/1.5003910} this term has been interpreted as Bohm contribution, which is not supported from a microscopic perspective. 

\subsection{Pressure tensors and linear response}
\label{Sec. IIc}

Linearizing around homogeneous equilibrium $n(\mathbf{r},t)=n_0+\delta n(\mathbf{r},t)$, $\mathbf{j}(\mathbf{r},t)=0+\delta \mathbf{j}(\mathbf{r},t)$ and $\hat{\Pi}(\mathbf{r},t)=\hat{\Pi}_0+\delta \hat{\Pi}(\mathbf{r},t)$ in the QHD [\ref{Eq_5a},\ref{Eq_5b}] and taking the space-time Fourier transform, $\mathfrak{F}$, an \emph{exact} equation for the functional derivative of the pressure can be found:
\begin{equation}\label{Eq_7}
\mathfrak{F}\left(\frac{\delta \hat{\Pi}^{kk}(r,t)}{\delta n(r',t')}\right)=\frac{\delta \Pi^{kk}}{\delta n}(\omega,k)=\frac{\omega^2}{k^2}-\frac{n_0}{m\chi(\omega,k)}
\end{equation}
where $\chi(r,r',\omega)=\delta n/\delta v_{eff}$ is the linear response (polarization) function (LRF) and the superscript $kk$ indicates the $\mathbf{e_k}\times\mathbf{e_k}$ component of the tensor. For the fermionic gas, $\chi$ can be analytically computed as Lindhard function\cite{9780521527965}:
\begin{align*}
\nonumber&\chi^L(\omega,k)=\pm\frac{dn_0}{2E_F}\frac{k_F}{k}\Psi_d(\frac{\omega+i\eta}{k v_F}\mp\frac{k}{2k_F})\\
&\Psi_d(z)=\int_0^1dx x^{d-1}\int d\Omega_d/\Omega_d(z-x cos(\theta))^{-1}\nonumber\\
&\Psi_3(z)=\frac{z}{2} + \frac{1 - z^2}{4}\log\frac{z+1}{z-1}.\nonumber
\end{align*}
Eq. \ref{Eq_7} asserts for any functional $\hat{\Pi}[n]$ a LRF $\chi^{\Pi}$. In Table \ref{table1} are compared the asymptotic $\chi^L$ and the LRF associated with $\Pi^0_{\alpha,\lambda}$. By direct comparison, the coefficients found in \cite{doi:10.1063/1.5003910} can be seen to be valid. But this recipe with spectrally dependent prefactors cannot be true since it was shown in Section \ref{Sec. IIb} that $\lambda=1$ universally. 
\begin{center}\label{table1}
	\begin{tabular}{ |c|c|c|c| } 
		\hline
		& $A$ & $\chi^{\alpha,\lambda}/A$ & $\chi^{L}/A$ \\ 
		\hline
		$\omega=0, k\ll k_F$ & $-\frac{3mn_0}{\hbar^2k_F^2}$ & $\frac{1}{\alpha}-\frac{3\lambda k^2}{4\alpha^2k_F^2}$ & $1-\frac{\lambda k^2}{12k_F^2}$ \\ 
		\hline
		$\omega=0, k\gg k_F$ & $-\frac{4mn_0}{\hbar^2k^2}$ & $\frac{1}{\lambda}-\frac{4\alpha k_F^2}{3k^2\lambda^2}$ & $1+\frac{4k_F^2}{5k^2}$  \\ 
		\hline
		$\omega\to\infty$ & $\frac{k^2n_0}{m\omega^2}$ & $1+\frac{\alpha k^2 k_F^2 \hbar ^2}{3 m^2 \omega ^2}+ \frac{\lambda k^4 \hbar ^2}{4 m^2 \omega ^2}$ & $1+\frac{3k^2 k_F^2 \hbar ^2}{5 m^2 \omega ^2}+\frac{ k^4 \hbar ^2}{4 m^2 \omega ^2}$ \\ 
		\hline
	\end{tabular}
	\captionof{table}{Asymptotic expansions for the Lindhard function and the LRF reproduced with $\hat{\Pi}^0_{\alpha,\lambda}$ functionals.}
\end{center}

\subsection{A nonlocal functional}
\label{Sec. IId}

The standard philosophy of Orbital-Free-DFT is based on the Local Density Approximation (LDA): a functional which reproduces exactly the pressure (through a set of constrains) for a specific system (usually homogeneous) is found and its closed form is generalized to any density $n(\mathbf{r},t)$ (the LDA limit). The present functional will be derived imposing the before mentioned equilibrium and linear dynamics constrains:
\begin{align}\label{Cond1}&\hat{\Pi}[n_0]=P_{TF}[n_0]\hat{1}\hspace{1cm}  n_0=const, \\ 
&\frac{\delta \Pi^{kk}}{\delta n}(\omega,k)=\frac{\omega^2}{k^2}-\frac{n_0}{m\chi^L(\omega,k)}.\label{Cond2}
\end{align}
In Section \ref{Sec. IIb} it has been suggested based on microscopic considerations that any KPDF can be decomposed in a zero order macroscopic Bohm and a first order, fermionic, macroscopic TF pressure along with another unknown term. This decomposition is proven in detail from a kinetic perspective in Appendix \ref{Sec. Va}. Taking into account the analytic form of Lindhard function in the Fourier domain for the condition \ref{Cond2}, it becomes obvious that $\delta \hat{\Pi}(r,t)/\delta n(r',t')$  must be non-local in space and time. The locality of the TF-Bohm terms implies time-nonlocality of the residual term $\hat{\Pi}_{NL}[n]$. By dimensional analysis (Appendix \ref{Sec. Va}):

\begin{align*}
&\hat{\Pi}[n]=-\frac{\hbar^2}{4m^2}n\nabla\otimes\nabla \ln n+P_{TF}[n]\hat{1}+\hat{\Pi}_{NL}[n]\nonumber\\
&\hat{\Pi}_{NL}=\frac{\hbar^2}{2m^2}\int dx'\delta(x-x')(\nabla_{\mathbf{r}}\otimes\nabla_{\mathbf{r}'}+\nabla_{\mathbf{r'}}\otimes\nabla_{\mathbf{r}})\tilde{\rho}(x,x')\\
&\tilde{\rho}(x,x')=\int dy dy'n^{1/2}(y)\mathcal{O}(x,y,x',y';n(x),n(x'))n^{1/2}(y)
\end{align*}
where $x=(\mathbf{r},t)$, $y=(\mathbf{r'},t')$ and the double density dependent kernel $\mathcal{O}(x,y,x',y';n_1,n_2)$ has been introduced. Further detailed calculus is presented in Appendix \ref{Sec. Va}. Due to translational invariance of the ground state density $n_0=constant$ the kernel is assumed invariant $\mathcal{O}(x,y,x',y;n_0,n_0)\equiv\mathcal{O}(x-y,x'-y';n_0,n_0)$ and the constrains [\ref{Cond1},\ref{Cond2}] are worked out within a Fourier representation $\mathcal{O}(\xi,\zeta;n_1,n_2)$ with $\xi\equiv(\omega,\mathbf{k})$. The outcome, together with a supplementary anzatz and the LDA limit:
$$\mathcal{O}(x,y,x',y';n(x),n(x'))\equiv \lim\limits_{\substack{n_1\to n(x)\\ n_2\to n(x')}}\mathcal{O}(x,y,x',y';n_1,n_2)$$
lead us towards the central result of this work:
\begin{align}\label{Eq_8}
&\hat{\Pi}_{NL}[n]=\int dx'\mathcal{L}(x,x')[n^{1/2}(x)\mathcal{D}(x')+n^{1/2}(x')\mathcal{D}(x)]\nonumber\\
&\mathcal{L}(x,x')=\delta(x-x')(\nabla_{\mathbf{r}}\otimes\nabla_{\mathbf{r}'}+\nabla_{\mathbf{r'}}\otimes\nabla_{\mathbf{r}})(\nabla_{\mathbf{r}}\nabla_{\mathbf{r'}})^{-1}\nonumber\\
&\mathcal{D}(x)=\frac{1}{2}\int dy \int d\xi e^{-i\xi(x-y)}\phi(\xi;n(x))n^{1/2}(y)\nonumber\\
&\phi(\xi;n_0)=\frac{\omega^2}{k^2}-\frac{n_0}{m\chi(\omega,k)}-\left(\frac{\delta \Pi_B}{\delta n}+\frac{\delta \Pi_{TF}}{\delta n}\right)_{n_0}.
\end{align}

The functional [\ref{Eq_8}] presents two levels of complexity. First, the operator $\mathcal{L}(x,x')$ involves solving an intricate $6D$ partial differential equation. This can be removed either considering various symmetries of the system or using the free energy density as trace of the pressure tensor: $\tau_{NL}[n]=Tr\Pi_{NL}[n]$:
\begin{align*}
\hat{\Pi}_{NL}\approx\tau_{NL}[n]\nabla\ln n\otimes\nabla\ln n\\
\nabla\cdot\hat{\Pi}_{NL}\approx\nabla\tau_{NL}[n].
\end{align*}

Both choices are consistent with the conditions [\ref{Cond1},\ref{Cond2}]. The second difficulty is related to the time-nonlocality which requires convoluting the kernel $\phi$ with the density at all times. The convolution is partially removed by the causal character of the LRF $\chi^L$ which makes $\mathcal{D}$ causal:

\begin{equation}\label{Eq_9}
\mathcal{D}(\mathbf{r},t)=\frac{1}{2}\int d\mathbf{r'}\int_{-\infty}^tdt'\phi(|\mathbf{r-r'}|,|t-t'|;n(\mathbf{r},t))n(\mathbf{r'},t')^{1/2}.
\end{equation}
Even though this form can be implemented in principle, it represents a tremendously difficult numerically workload. A workaround this problem is further presented.

\subsection{An approximative functional}
\label{Sec. IIe}

The formula \ref{Eq_9} suggests that $\mathcal{\phi}(|\mathbf{r-r'}|,|t-t'|;n(\mathbf{r},t))$ can be seen as a propagator, i.e. a Green function which in turn, could be given by an integral equation for $\mathcal{D}$. On the other hand, local (differential) equations have Green functions which in the Fourier representation can be expressed as rational functions. Motivated by this idea and inspired by Drude-like approximations of the dielectric constant together with the asymptotic behavior of the kernel $\phi$, the following approximative form is proposed:
\begin{align}\label{approx}
\phi^{app}(\omega,k)\approx \phi_0^\infty\frac{\omega^2-i\gamma(k) \omega+t_2(k)}{\omega^2-i\gamma(k)\omega+t_1(k)}
\end{align}
designed to reproduce the kernel $\phi(\omega,k)$ exactly at the asymptotic limits $\omega=0$, $\omega\to\infty$ and the mid-line inside the particle-hole continuum, i.e. $\omega=\hbar^2k^2/2m+\hbar k/m$. The presence of the imaginary term has a three-fold importance: it helps reproduce the exact kernel, models the dissipative phenomena (Landau damping) and makes the kernel analytical in the complex plane. The last property implies causality and, consequently, the validity of the F-sume rule:
$$-\frac{2}{\pi}\int_{-\infty}^{\infty}\omega \mathcal{I}m[\chi^{app}](\omega,q)=\frac{n_0q^2}{m}$$
Skipping the details of the calculus presented in Appendix \ref{Sec. Vb}, the approximation [\ref{approx}] allows us to write a wave-like equations for $\mathcal{D}$ with non-constant coefficients and source:
\begin{eqnarray}\label{Eq_10}
[\partial_{t,t}-\hat{\gamma}\partial_t-\hat{t}_1]\mathcal{D}=\frac{4\hbar^2k_F^2}{15m^2} [\partial_{t,t}-\hat{\gamma}\partial_t-\hat{t}_2]n^{1/2}
\end{eqnarray}
where the operators $\hat{\gamma}[n]$, $\hat{t}_1[n]$ and $\hat{t}_2[n]$ and their action on spatial functions are defined in Appendix \ref{Sec. Vb}. The term $\mathcal{D}$ is now an approximation of the \emph{exact} one prescribed by Eq. [\ref{Eq_9}], but the one which should be used in practice as it is much more easily to compute numerically thanks to its local-in-time nature.

Through the identity $\tau[n]=Tr\hat{\Pi}[n]$ one can develop a field-theoretical QHD \cite{doi:10.1063/1.5003910,doi:10.1002/ctpp.201500024,0953-4075-48-18-185102} where the approximation $\nabla\cdot\hat{\Pi}[n]=n\nabla (\delta T[n]/\delta n)$ holds and the kinetic density functional:
$$T[n]=2\int dxdx'\delta(t-t')n(x)^{1/2}\phi(x-x';n(x))n(x')^{1/2}.$$

At this end, let us collect the full prescription of the functional (from now on we shall refer to as KPDF) which will be always used in practice instead of the one given in Eq. [\ref{Eq_9}] as it is orders of magnitude easier to be numerically implemented:
\begin{align*}
&\hat{\Pi}[n]=\frac{\hbar^2nk_F^2}{5m^2}\hat{1}-\frac{\hbar^2}{4m^2}n\nabla\otimes\nabla\ln n+\hat{\Pi}_{NL}[n]\\
&\hat{\Pi}_{NL}[n]=\int dx'\mathcal{L}(x,x')[n^{1/2}(x)\mathcal{D}(x')+n^{1/2}(x')\mathcal{D}(x)]\\
&\mathcal{L}(x,x')=\delta(x-x')(\nabla_{\mathbf{r}}\otimes\nabla_{\mathbf{r}'}+\nabla_{\mathbf{r'}}\otimes\nabla_{\mathbf{r}})(\nabla_{\mathbf{r}}\nabla_{\mathbf{r'}})^{-1}\\
&(\partial_{t,t}-\hat{\gamma}\partial_t-\hat{t}_1)\mathcal{D}=\frac{4\hbar^2k_F^2}{15m^2}(\partial_{t,t}-\hat{\gamma}\partial_t-\hat{t}_2)n^{1/2}\\
&\nabla\cdot\hat{\Pi}_{NL}[n]\approx 4\nabla (n^{1/2}\mathcal{D}).
\end{align*}

\section{Results}
\label{Sec. III}
\subsection{Accuracy of the approximative kernel}
\label{Sec. IIIa}

Before testing the validity and the improvements brought by the KPDF for realistic systems, it is important to understand what is lost along the approximation [\ref{Eq_10}]. The best picture is provided by the comparison between the exact $\chi^L(w,q)$ and the $\chi^{app}(w,q)$ associated by Eqns. [\ref{Eq_7},\ref{approx}] with $\mathcal{D}$. From now on, the following scaling will be adopted: $w \equiv m\omega/\hbar k_F^2$, $q\equiv k/k_F$, $\chi\equiv\chi v_F^2$. 

\begin{figure}[H]
	\centering
	\subfloat{\includegraphics[width = 0.5\linewidth]{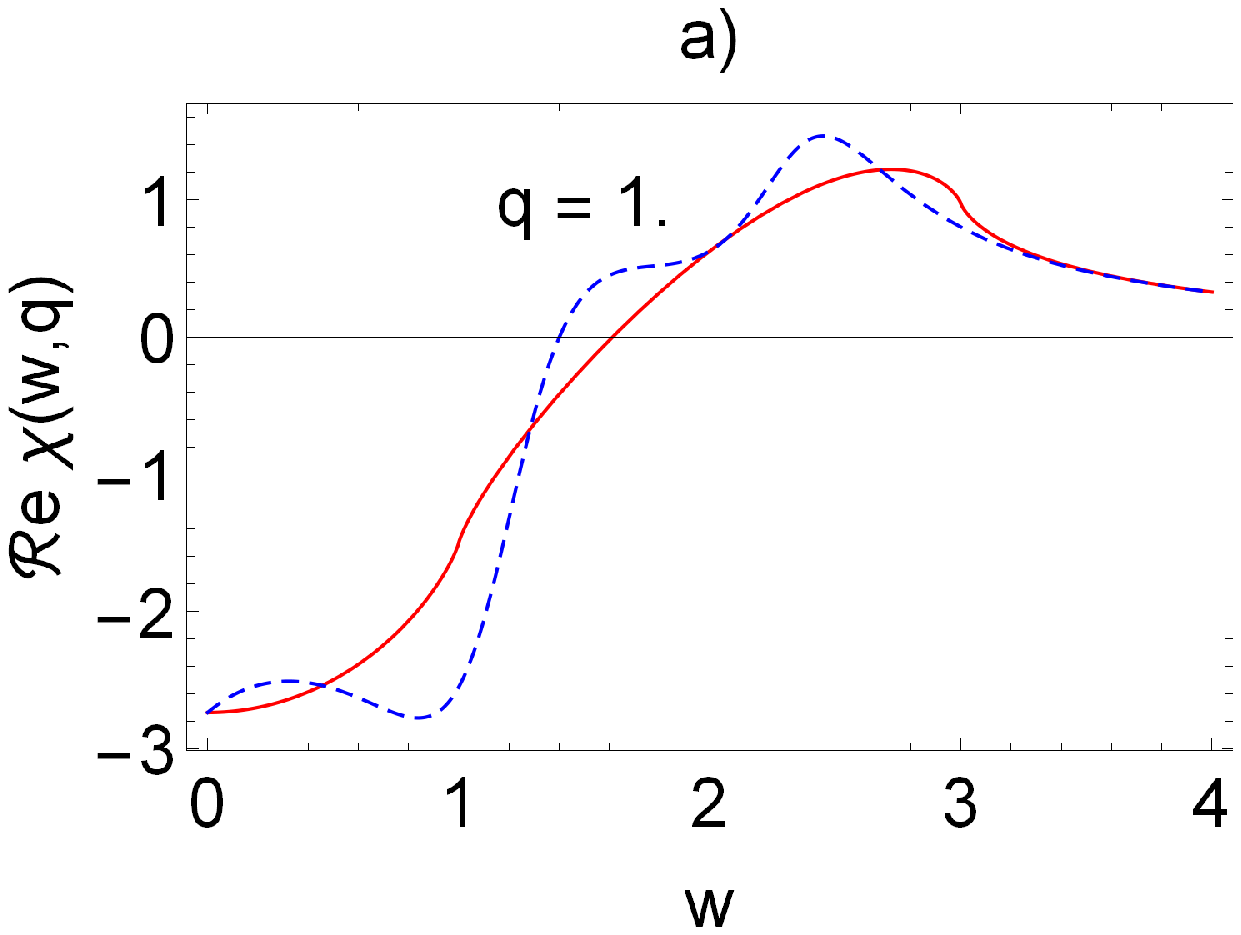}}
	\subfloat{\includegraphics[width = 0.5\linewidth]{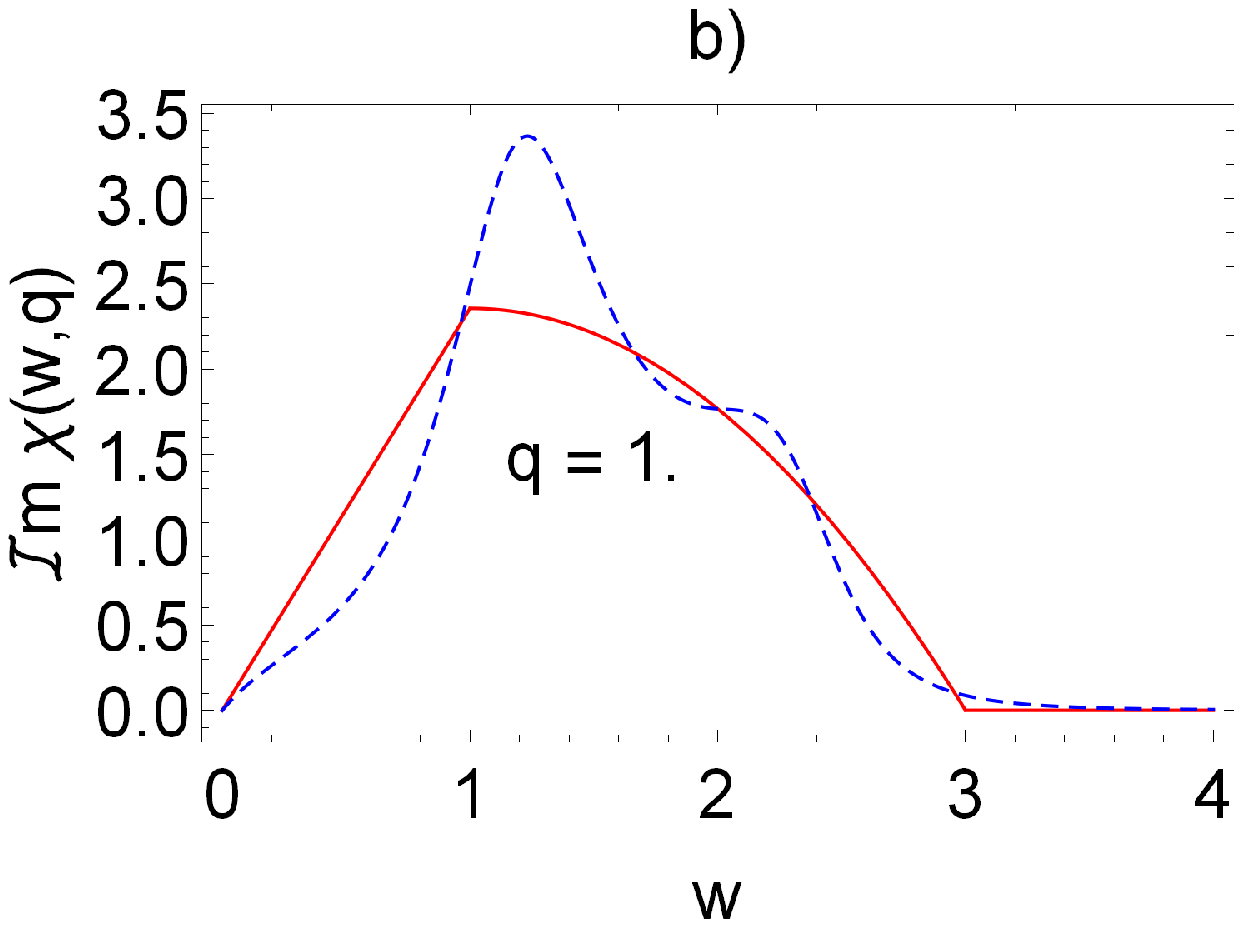}}\\
	\subfloat{\includegraphics[width = 0.5\linewidth]{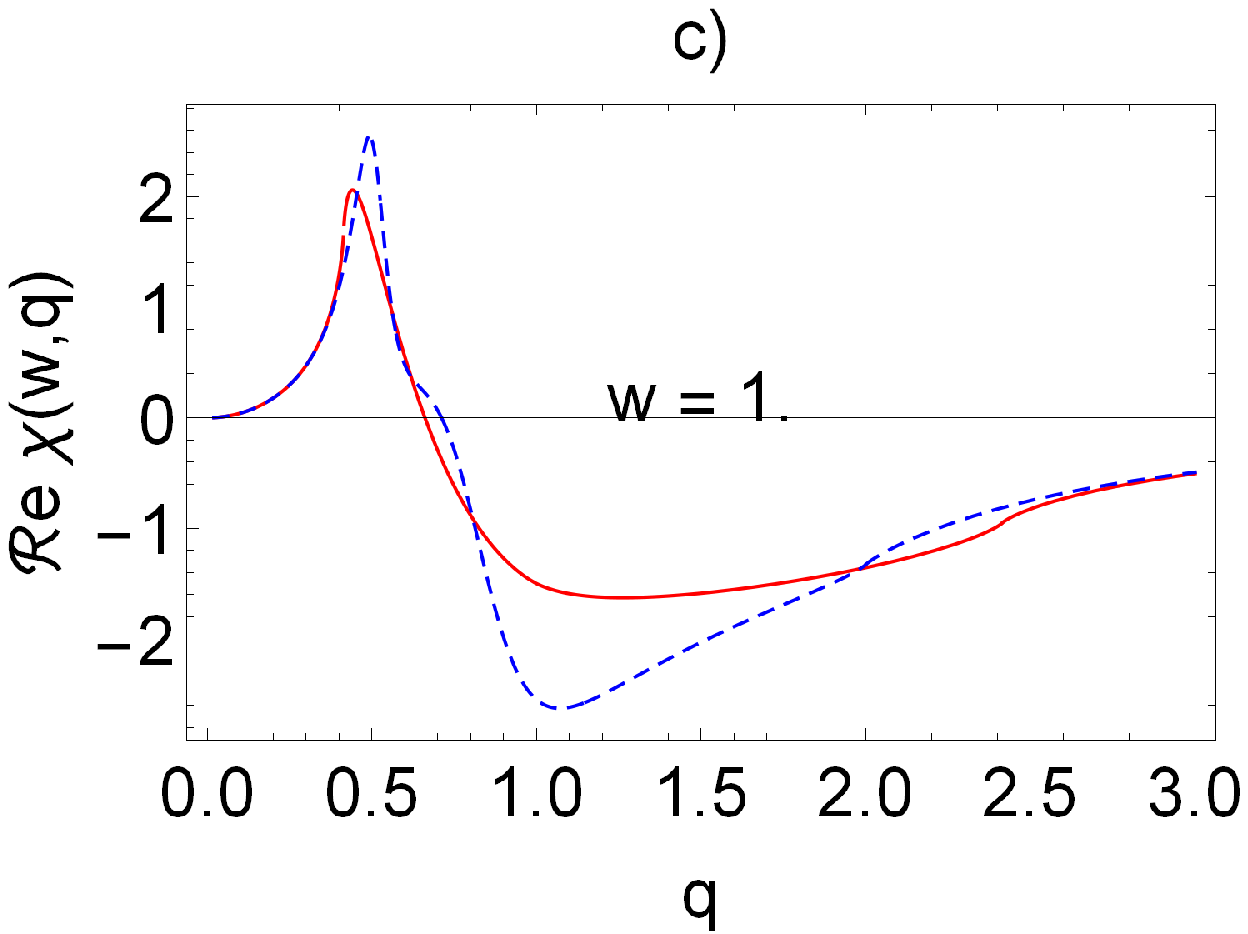}}
	\subfloat{\includegraphics[width = 0.5\linewidth]{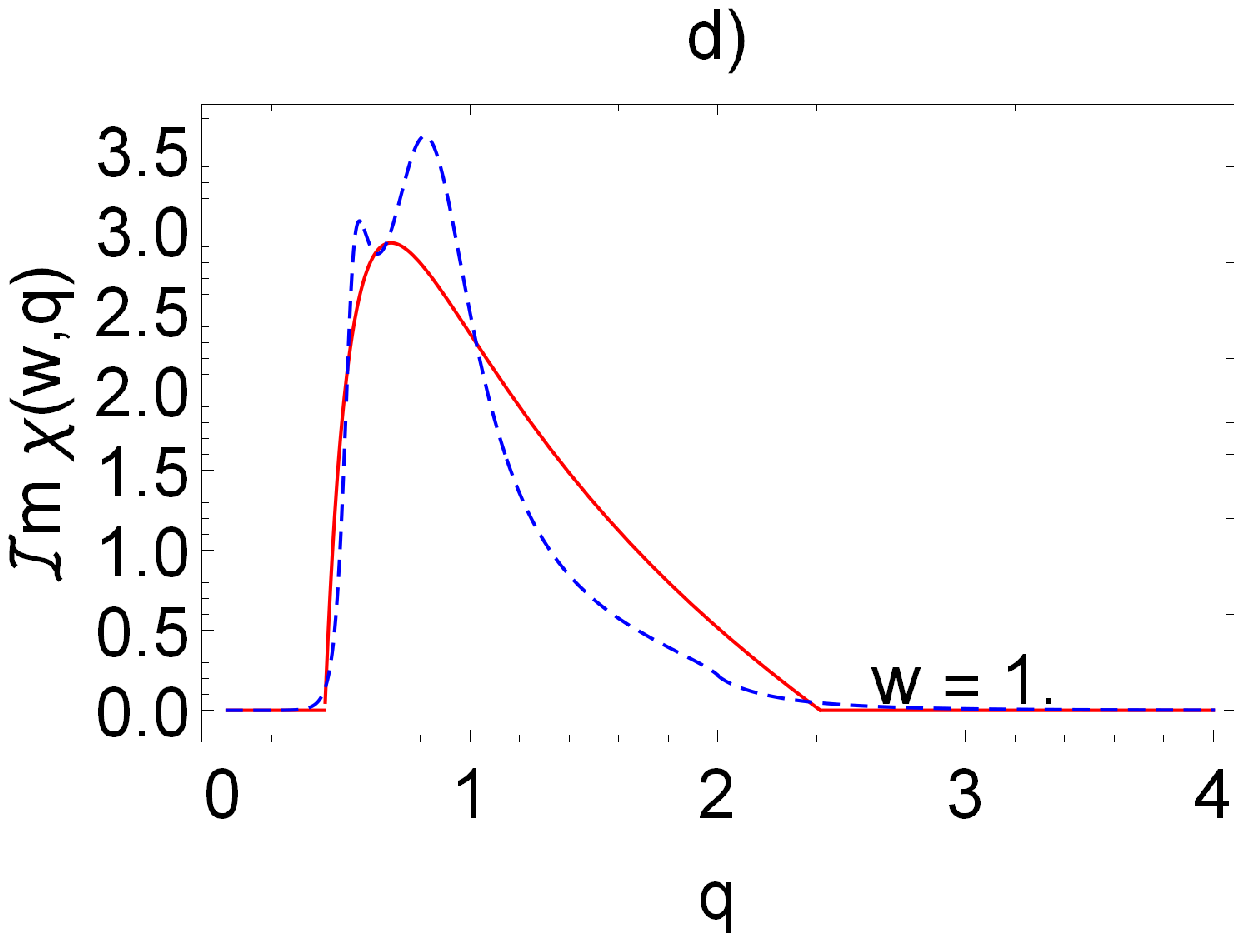}}
	\caption{Real ($a,c$) and imaginary ($b,d$) parts of the Lindhard function (red) and the approximative LRF $\chi^{app}$ (blue,dashed) at $q=1$ ($a, b$) and $w=1$ ($b, d$).}
	\label{Fig_1}
\end{figure}

Since, by design, the LRF's should agree well asymptotically ($w\gg q^2$ and $w\ll q^2$), in Fig. \ref{Fig_1} are plotted the real and imaginary parts of $\chi^L(w,q)$  and $\chi^{app}(w,q)$ at $w=1$ and $q=1$, where large discrepancies are expected. The exact profile of $\chi^L$ is fairly well interpolated by $\chi^{app}$ in between the asymptotics. As a pitfall, a smooth tail appears in the imaginary part of $\chi^{app}(w,q)$ outside the particle-hole continuum, indicating a pathological presence of the damping. This behavior is a consequence of the smooth analytic form \ref{approx} which cannot reproduce the logarithmic discontinuities of $\chi^{L}$ (its derivatives).

\begin{figure}[H]
	\centering
	\subfloat{\includegraphics[width = 0.5\linewidth]{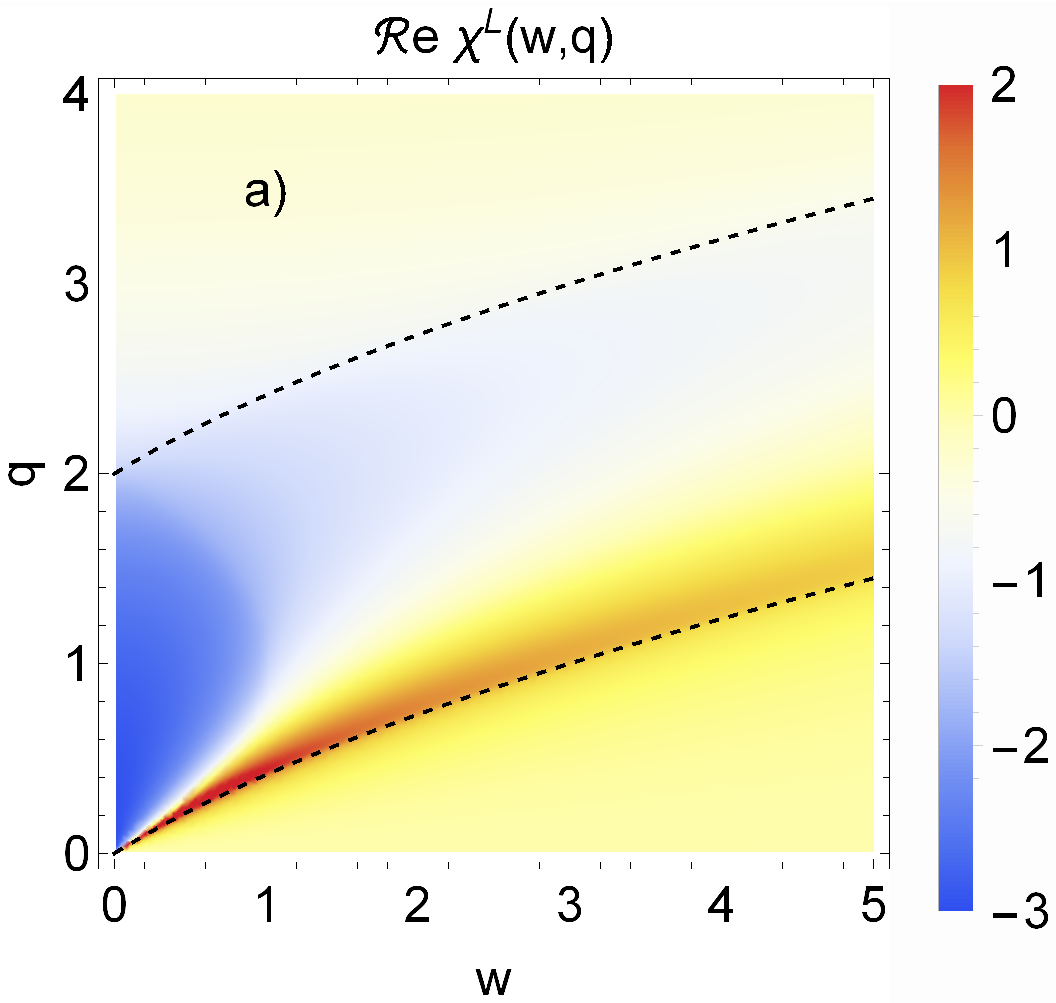}}
	\subfloat{\includegraphics[width = 0.5\linewidth]{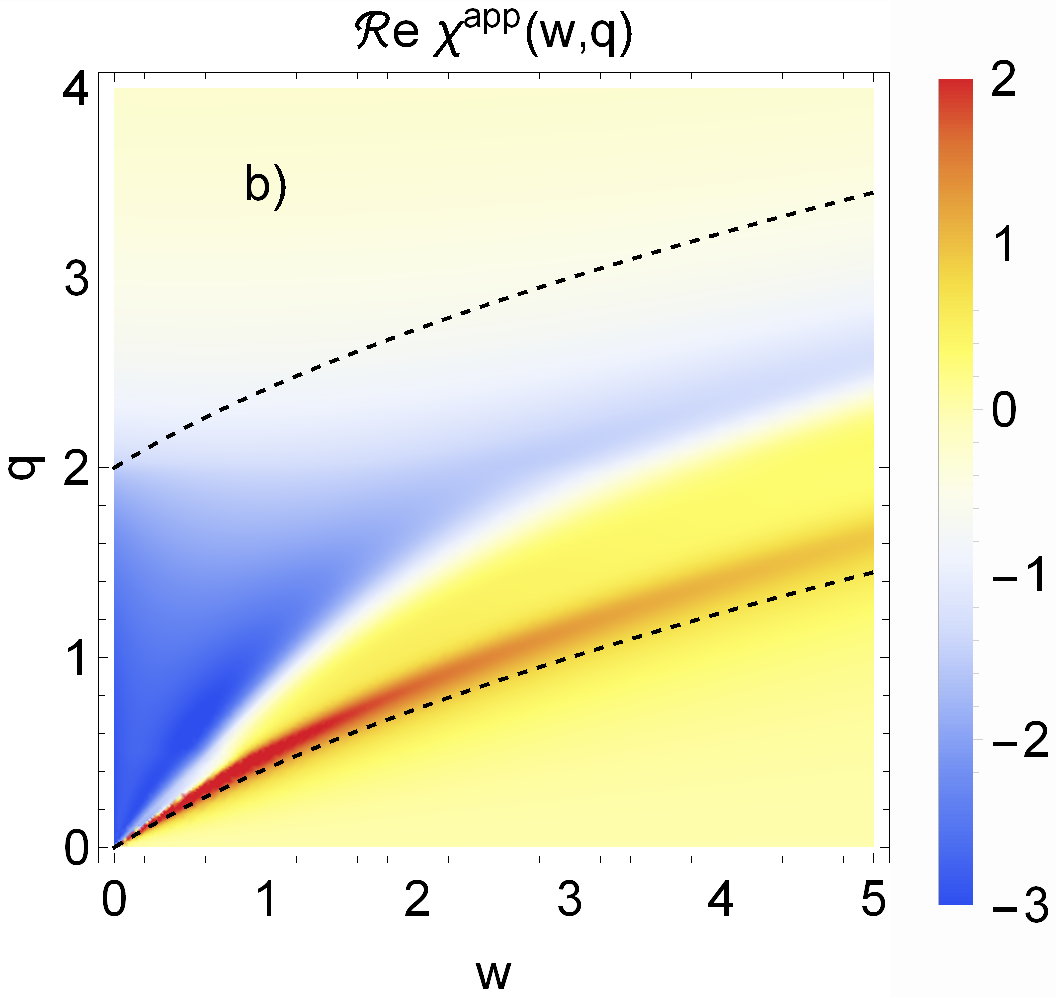}}\\
	\subfloat{\includegraphics[width = 0.5\linewidth]{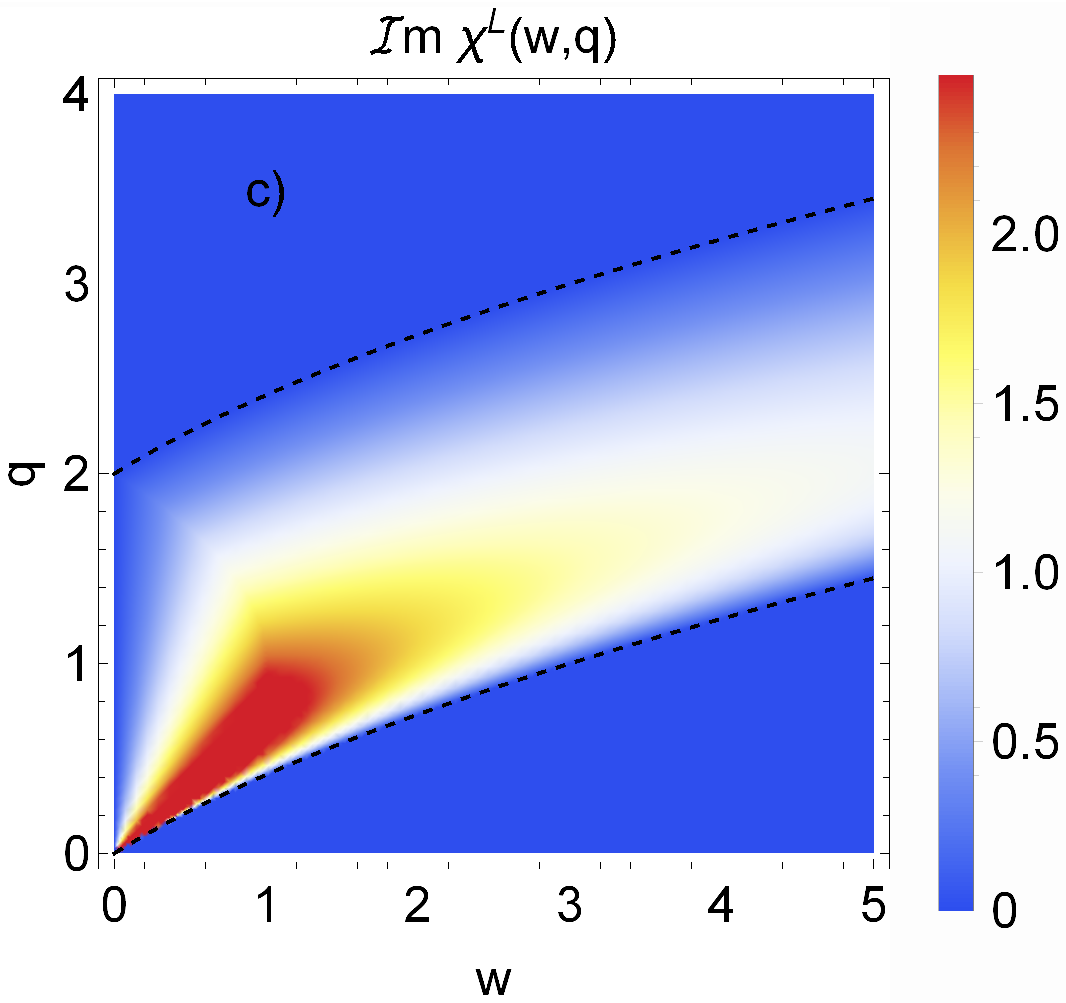}}
	\subfloat{\includegraphics[width = 0.5\linewidth]{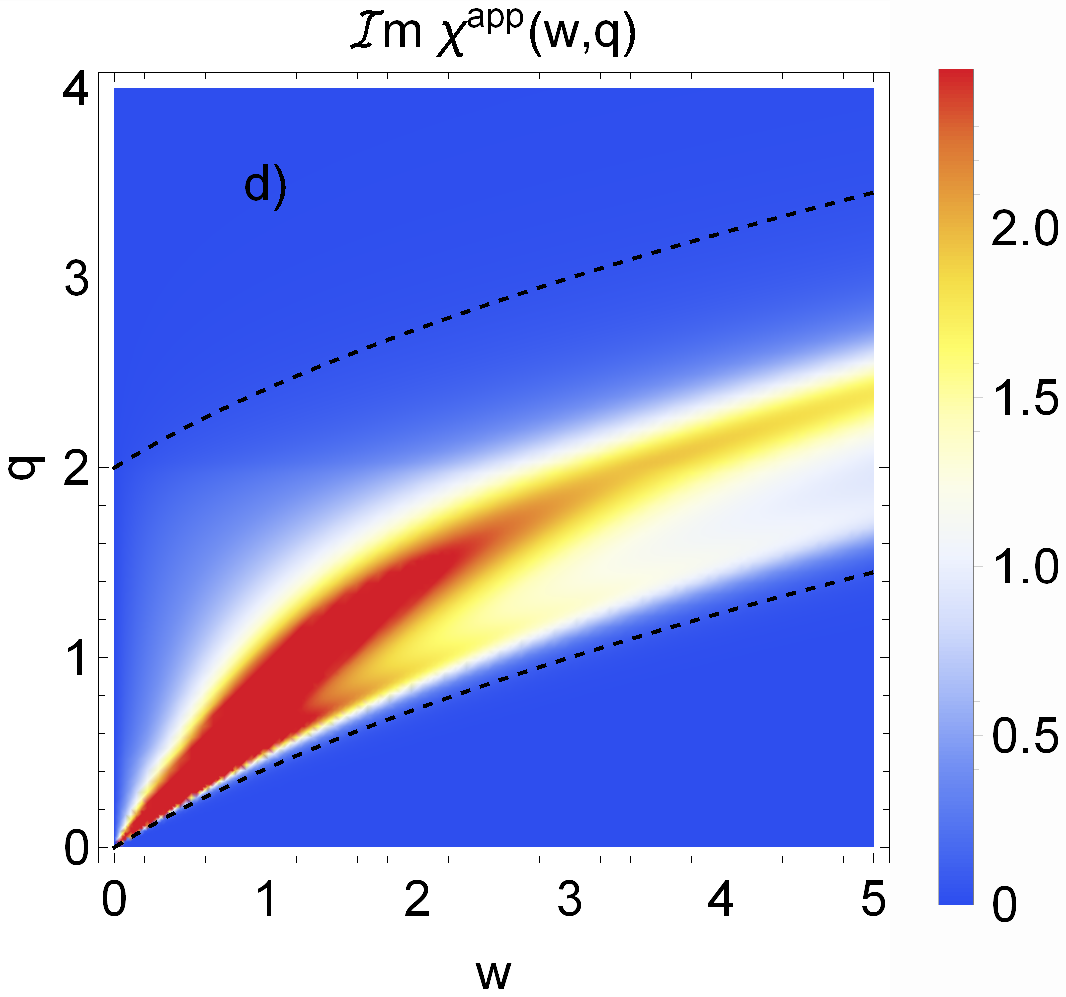}}
	\caption{Real ($a,b$) and imaginary ($c,d$) parts of the Lindhard function ($a,c$) and the approximative LRF $\chi^{app}$ ($b,d$).}
	\label{Fig_2}
\end{figure}

A comprehensive comparison is shown in Fig. \ref{Fig_2} for the same quantities as in Fig. \ref{Fig_1} but on the whole spectrum $(w,q)$ as density-plot.

As expected, the approximative KPDF works very well asymptotically, outside the particle-hole continuum defined by $w=max(q^2\pm 2q,0)$. While inside this region errors up to $50\%$ are expected, the trends of the Lindhard function are reproduced. This is expected to lead to good qualitative behavior in realistic simulations. Other approximations of the kernel $\phi$ compatible with time-locality and capable to reproduce even better the Lindhard function might be designed, but the present work is concerned with the simplest of them as it is the easiest to implement numerically.

\subsection{Extensive Monte Carlo tests}
\label{Sec. IIIb}

The results shown in Fig. \ref{Fig_1} - \ref{Fig_2} indicate the levels of accuracy for the approximative KPDF in the linear regime from a spectral perspective. However, in realistic scenarios, the external potential is induced as initial value quantity simultaneously affecting multiple spectral regions. Even more, one cannot estimate how the second order mode-coupling which is enabled beyond pure linear regime will be reflected in the KPDF. Finally, will the errors from the particle-hole continuum region have only a quantitative impact or also a qualitative one, perhaps leading to unphysical behaviors? To answer all these questions in an exhaustive manner a simple toy system is used.  

The system is a 3D Fermi gas under spatial periodic boundary conditions. For the stationary regime, the gas is assumed in its groundstate under the influence of static effective potential $v_0(z)$ while, for dynamics, the system is considered to be found in homogeneous ground state at $t<0$ and subjected to an effective potential $v(z,t)$ at $t\geq 0$. The unidirectional dependency on $Oz$ axes is chosen for simplicity, without spoiling the main conclusions of the analysis. The potential $v(z,t)$ supports a Fourier decomposition:
$$v(z,t)=\int d\omega dkA(\omega,k)e^{-i(\omega t-k z)}$$
Due to translational invariance in the $\mathbf{r}_\perp=(x,y)$ plane, the KS pseudo-orbitals for each particle can be represented as:
$$\Psi_{\mathbf{k}}(\mathbf{r}_\perp,z,t)=e^{-i\mathbf{k}_\perp \mathbf{r}_\perp}e^{-i\frac{\hbar\mathbf{k}_\perp^2}{2m}t}\psi_{k_z}(z,t)$$
$\mathbf{k}=\mathbf{k}_\perp+\hat{\mathbf{e}}_zk_z$ obeying KS Eqns:
\begin{equation}\label{Eq_11}
i\hbar\partial_t\psi_{k_z}(z,t)=\left[-\frac{\hbar^2}{2m}\partial_{zz}+v(z,t)\right]\psi_{k_z}(z,t)
\end{equation}
at homogeneous stationarity: $\psi_{k_z}(t=0)=\exp^{-ik_zz}$ where $k_F\le k_z\le k_F$. The macroscopic quantities can be obtained after averaging over the orthogonal degeneracy $\mathbf{k}_\perp$:
\begin{align*}
&n(z,t)=\frac{3}{4k_F^3}\int_{-k_F}^{k_F}(k_F^2-k_z^2)|\psi_{k_z}|^2dk_z\\
&j_z(z,t)=\frac{3\hbar}{4mk_F^3}\int_{-k_F}^{k_F}(k_F^2-k_z^2)Im(\psi_{k_z}^*\partial_z\psi_{k_z})dk_z\\
&\Pi_{\perp,\perp}(z,t)=\frac{3\hbar^2}{8m^2k_F^3}\int_{-k_F}^{k_F}(k_F^2-k_z^2)^2|\psi_{k_z}|^2dk_z\\
&\Pi_{z,z}(z,t)=\frac{3\hbar^2}{4m^2k_F^3}\int_{-k_F}^{k_F}(k_F^2-k_z^2)|\partial_z\psi_{k_z}|^2dk_z\\
\end{align*}
other quantities: $j_\perp=0$, $\Pi_{\perp,z}=\Pi_{z,\perp}=0$. 

The Eq. \ref{Eq_11} is solved numerically using a pseudo-spectral method \cite{9780898710236} on an uniform 1D grid. Variables and quantities are scaled as it follows: density by groundstate $n_0$, current by $v_F n_0$, pressure by $2n_0E_F/m$, the potential with $E_F$, the space variable $z$ with $k_F^{-1}$ and time by $(E_F/\hbar)^{-1}$. The spatial domain $L=\pi N$ is discretized in $2^{10}$ equidistant points, while $N=200$ was chosen to resolve the thermodynamic limit $N\to\infty$ of a Fermi gas. The temporal evolution is done via an operator-splitting technique with constant time-steps $\delta t\approx 10^{-3}$ to ensure reasonable ($\le 10^{-2}\%$) conservation of total norm, energy and momentum. Testing the approximative KPDF is equivalent to comparing $\Pi^{app}_{z,z}$ against the exact, microscopic $\Pi_{z,z}$, the former being computed from the \emph{exact} density profile obtained from microscopic simulations. 

\begin{figure}
	\centering
	\subfloat{\includegraphics[width = 0.9\linewidth]{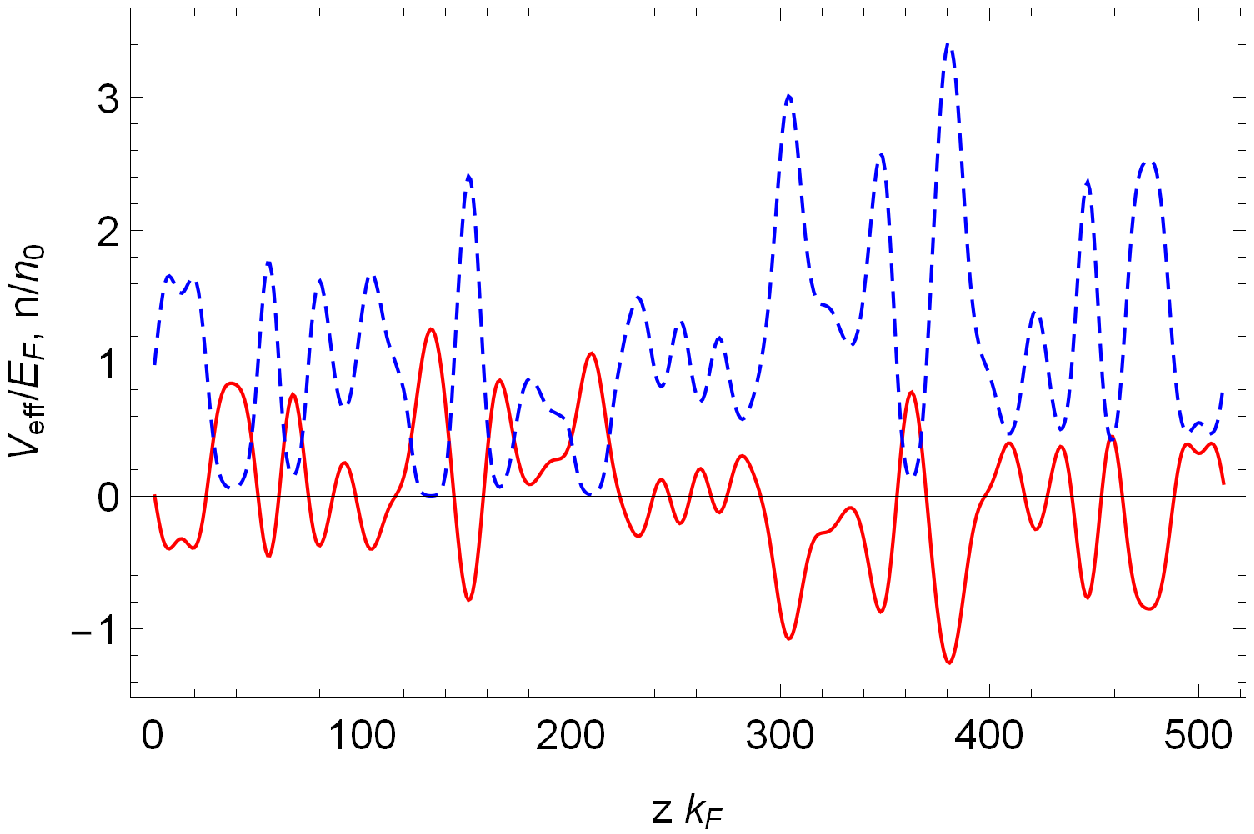}}\\
	\subfloat{\includegraphics[width = 0.9\linewidth]{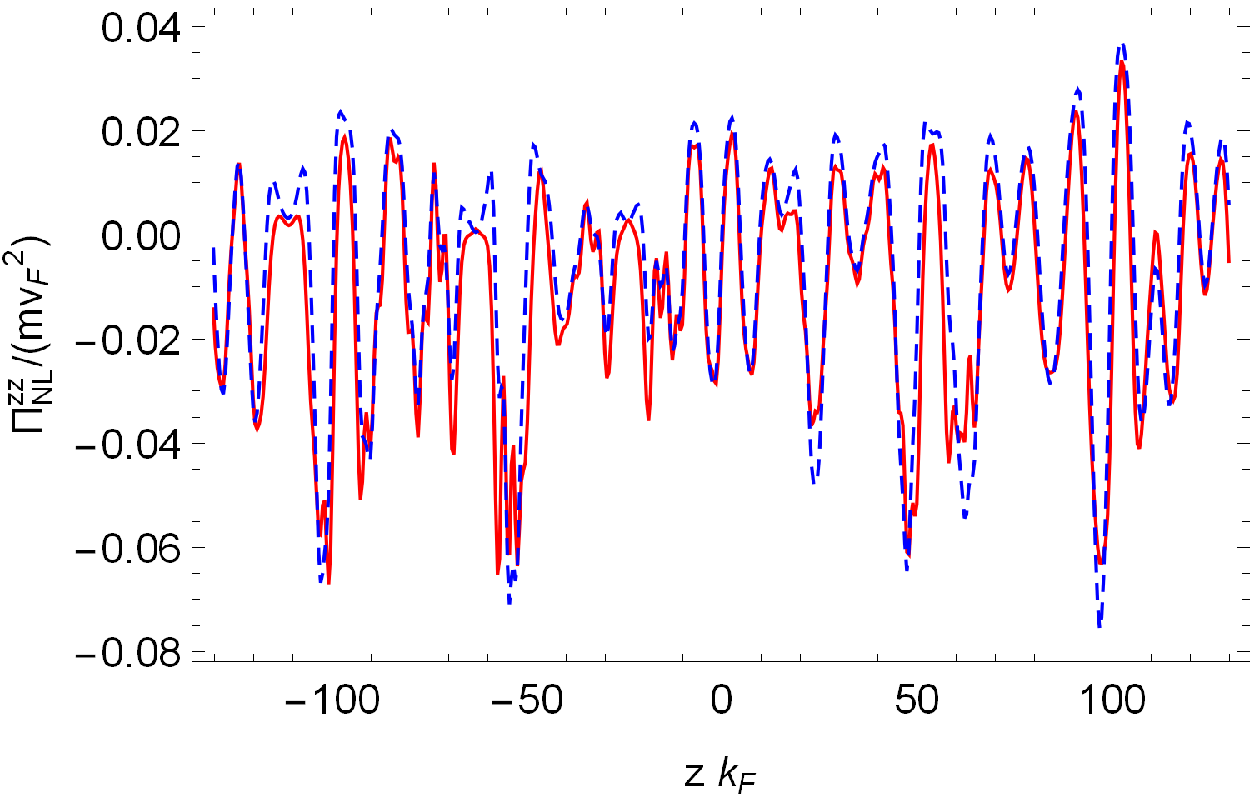}}
	\caption{Above: a generic random potential (red, full-line)  and the associated ground-state density profile of the Fermi gas (blue, dashed); Bellow: the spatial profile for the $z,z$ component of the non-local microscopic (red, full-line) and KPDF (blue, dashed) pressure.}
	\label{Fig_3}
\end{figure}

At this point the purpose is to test the KPDF in complicated scenarios, possibly beyond the linear regime, and compare it with the more recent functionals $\hat{\Pi}_{\alpha,\lambda}^0$ \cite{doi:10.1063/1.5003910}. This will be achieved considering a Monte Carlo ensemble of effective potentials, each one with a wide spectrum. The amplitudes $A(\omega,k)$ for each potential  $v(z,t)$ are randomly generated with a probability $P(A,w,q)\sim \exp[-5 |A|^2]\Theta(3-|w|)/(q^2+1)$. With this probability function, the system is forced to go beyond linear regime, on slow and fast time scales, at short and long wavelengths. 

\begin{figure}
	\centering
\includegraphics[width = 0.9\linewidth]{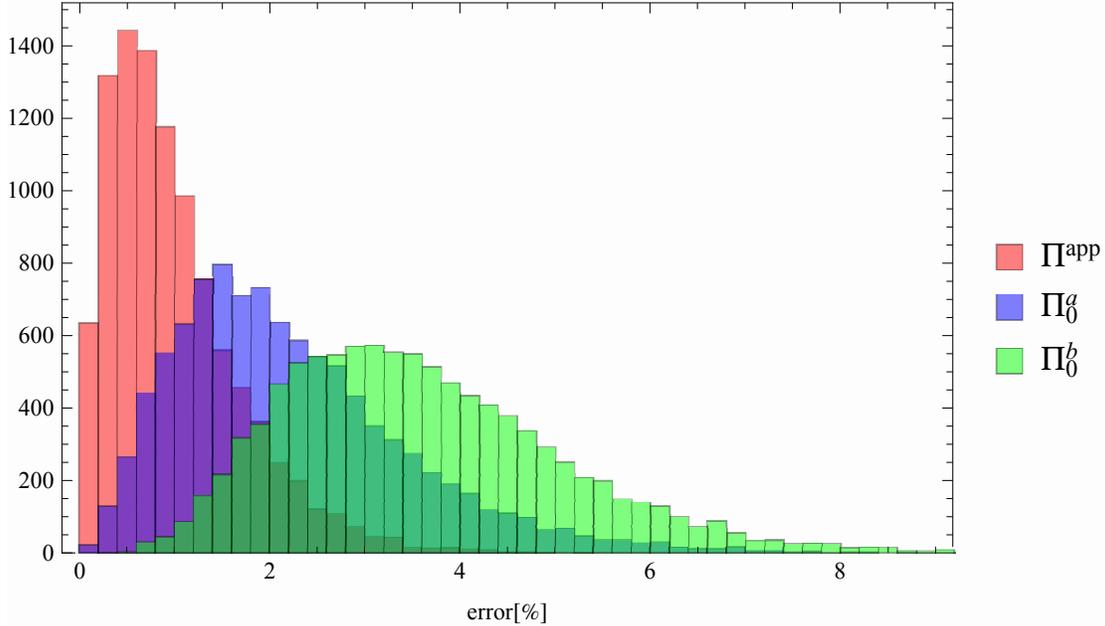}
	\caption{Distribution of error in the Monte-Carlo ensemble of stationary profiles provided by KPDF (red), $\hat{\Pi}_{1,1/9}^0$ (blue) and $\hat{\Pi}_{3/5,1}^0$ (green).}
	\label{Fig_4}
\end{figure}

\emph{Stationary states} are resolved solving the Eqns.[\ref{Eq_11}] by the imaginary time method. In Fig. [\ref{Fig_3}] are plotted the solutions of a generic potential (red, full line) from the ensemble and the associated total density (blue, dashed). It is found that for groundstates, even the TF-Bohm ($\lambda = 1$) provides good agreement. For that in the second figure, the comparison between non-local parts of the pressure profiles (beyond TF-Bohm) are plotted pointing out the power of the KPDF. Such results are generic for the whole ensemble, therefore we proceed to capture its statistics by computing the error in each case. The error is defined as space averaged local difference between microscopic and functional pressure $||\Pi_{z,z}-\Pi_{z,z}^{app}||$. In the histogram [\ref{Fig_4}] are shown the results for the approximative KPDF (red) in comparison with the ones provided by $\hat{\Pi}_{3/5,1}^0$ (green, valid at high wavenumbers) and by $\hat{\Pi}_{1,1/9}^0$ (blue, valid at low wave-numbers). Although it was not found any explanation for the gamma-like distributions, it is a clear representation of how, qualitatively, the KPDF is on average almost an order of magnitude more precise than $\hat{\Pi}_{\alpha,\lambda}^0$ approximations, yielding also a lower dispersion of the errors.

\begin{figure}
	\centering
	\subfloat{\includegraphics[width = 0.45\linewidth]{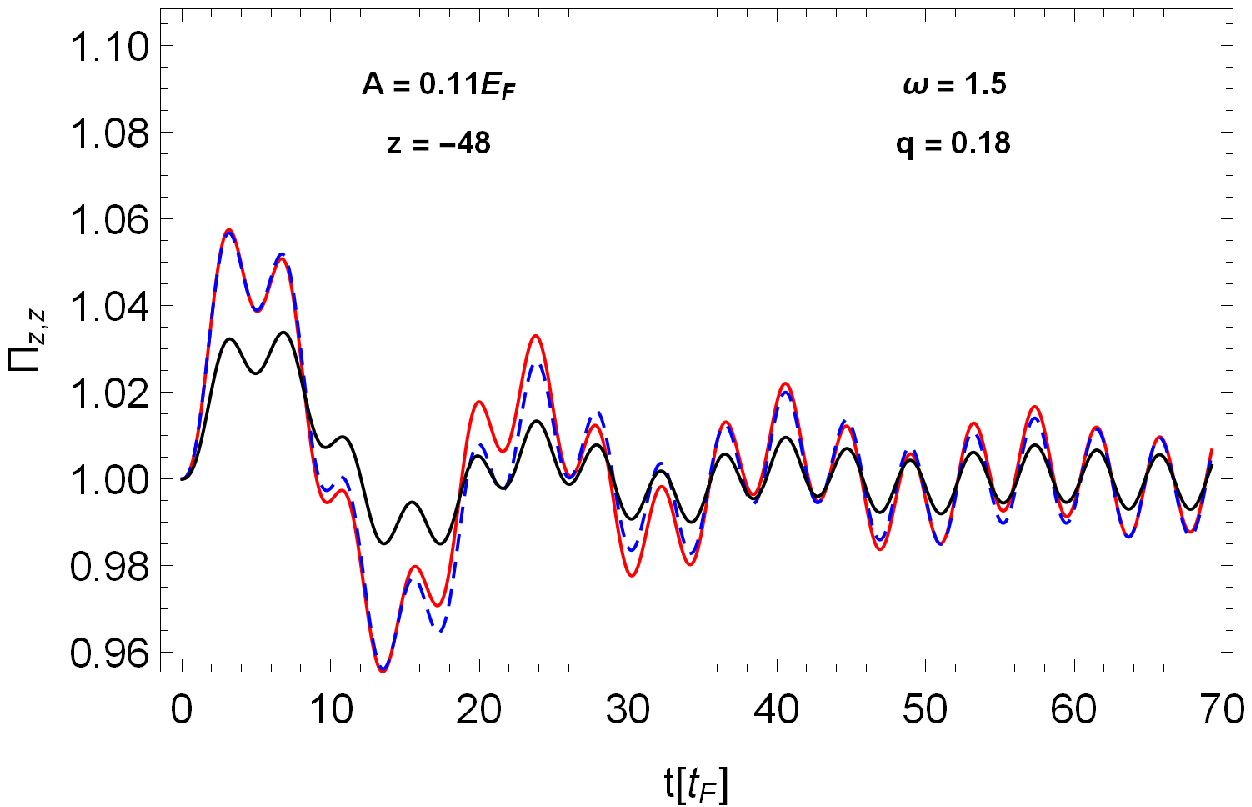}}
	\subfloat{\includegraphics[width = 0.45\linewidth]{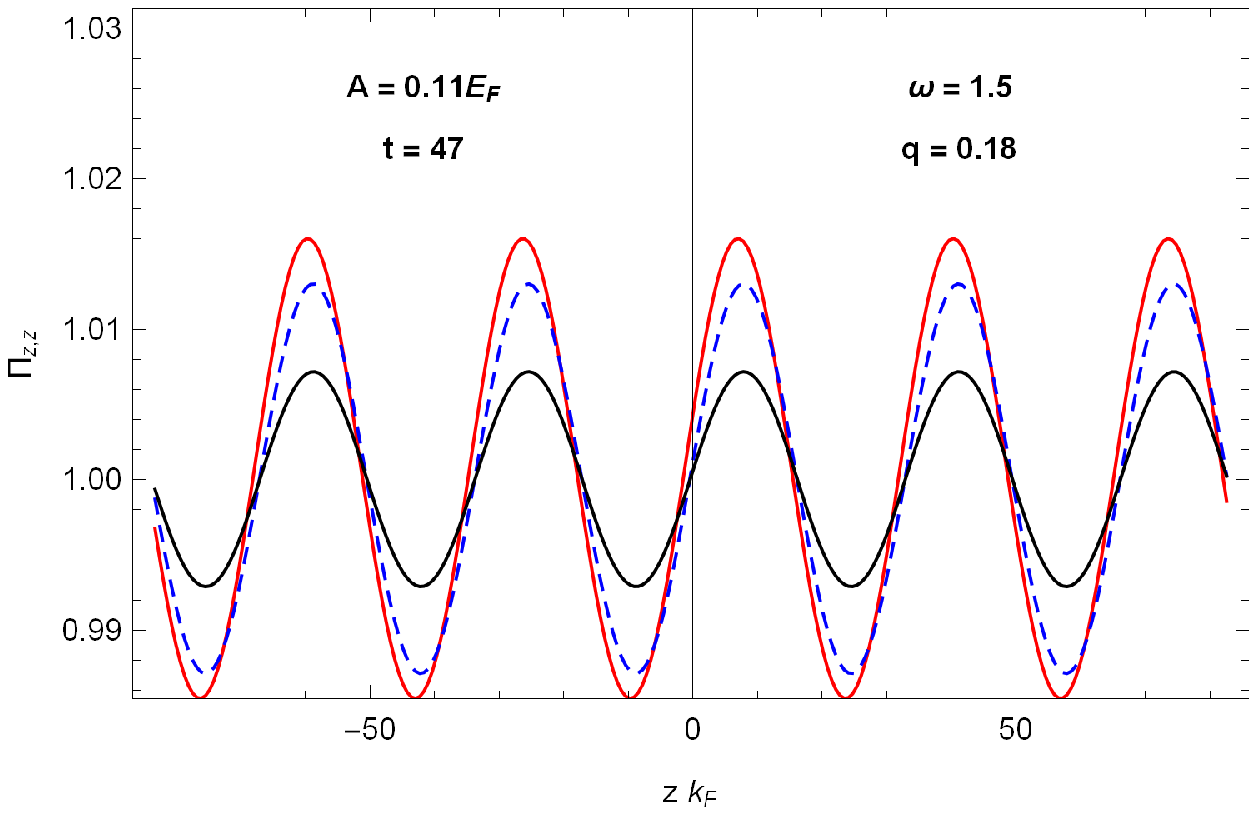}}\\
	\subfloat{\includegraphics[width = 0.45\linewidth]{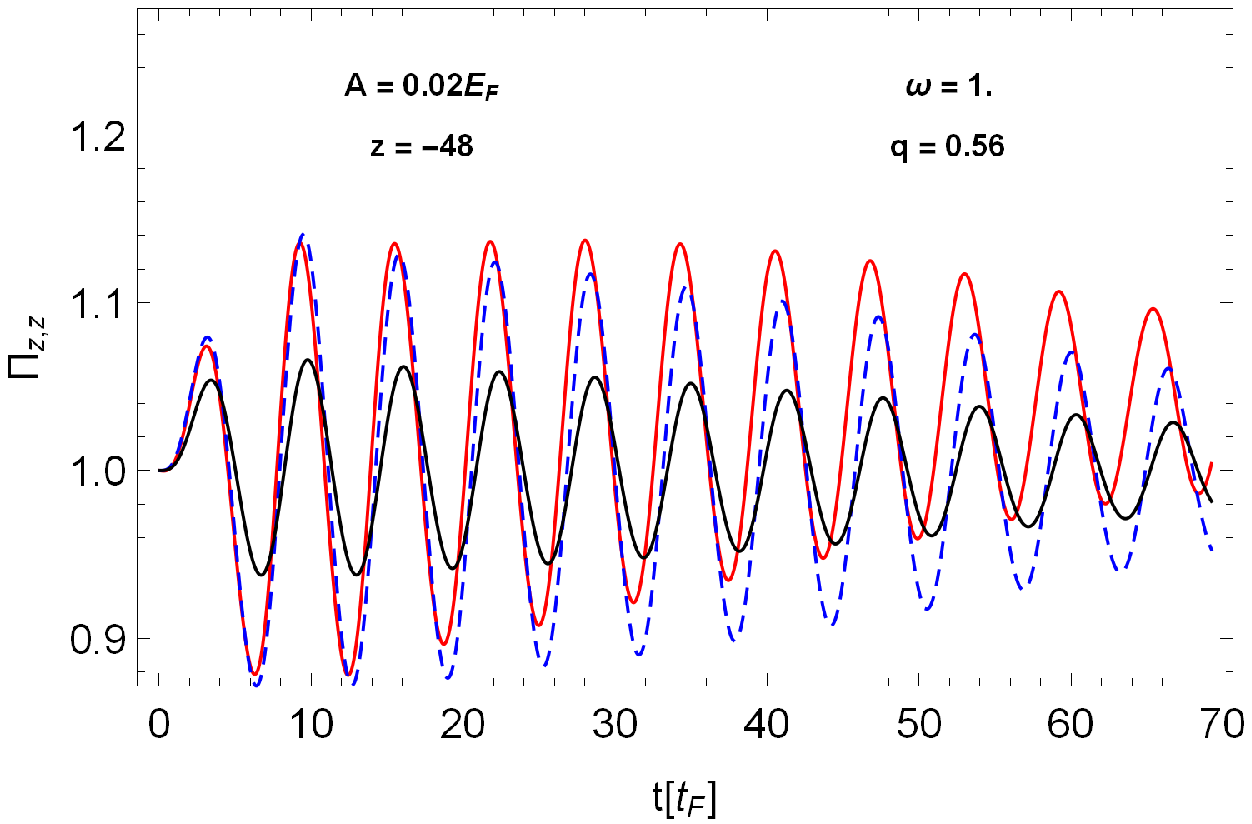}}
	\subfloat{\includegraphics[width = 0.45\linewidth]{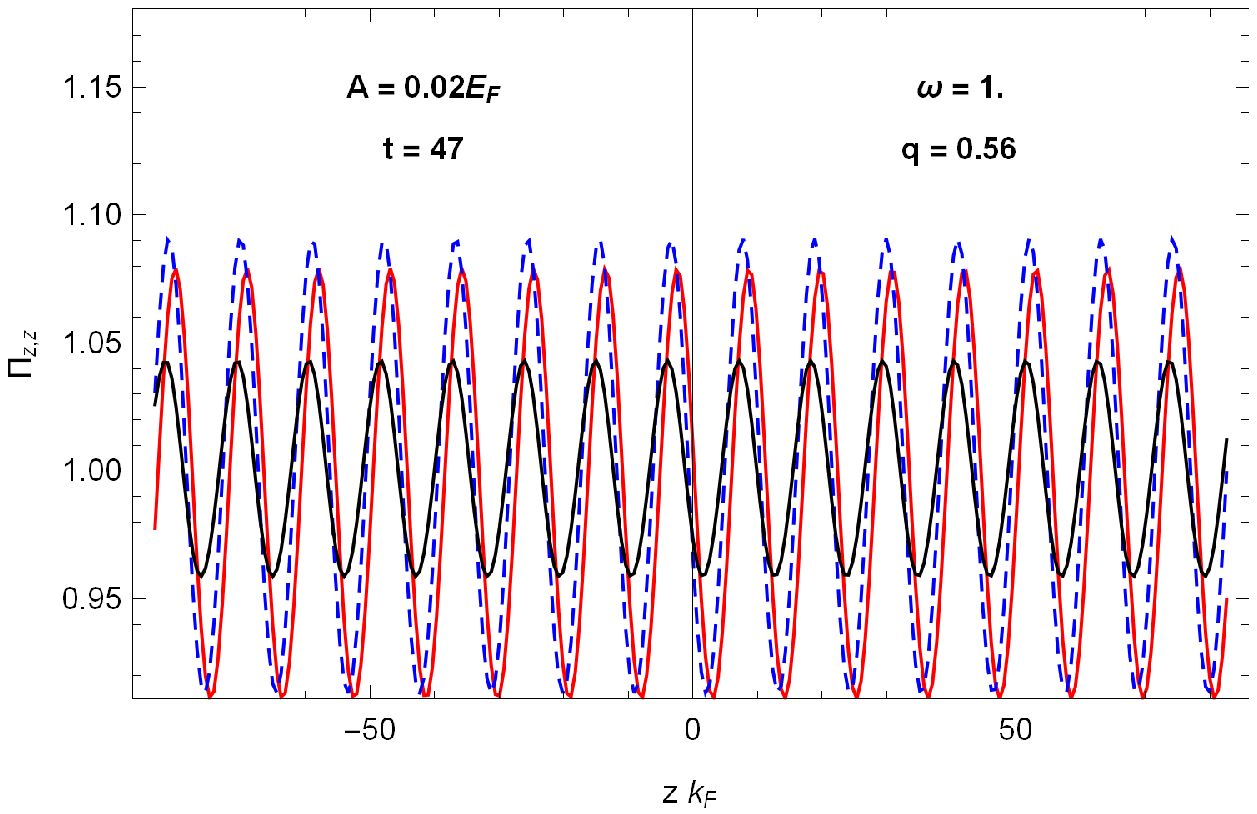}}\\
	\subfloat{\includegraphics[width = 0.45\linewidth]{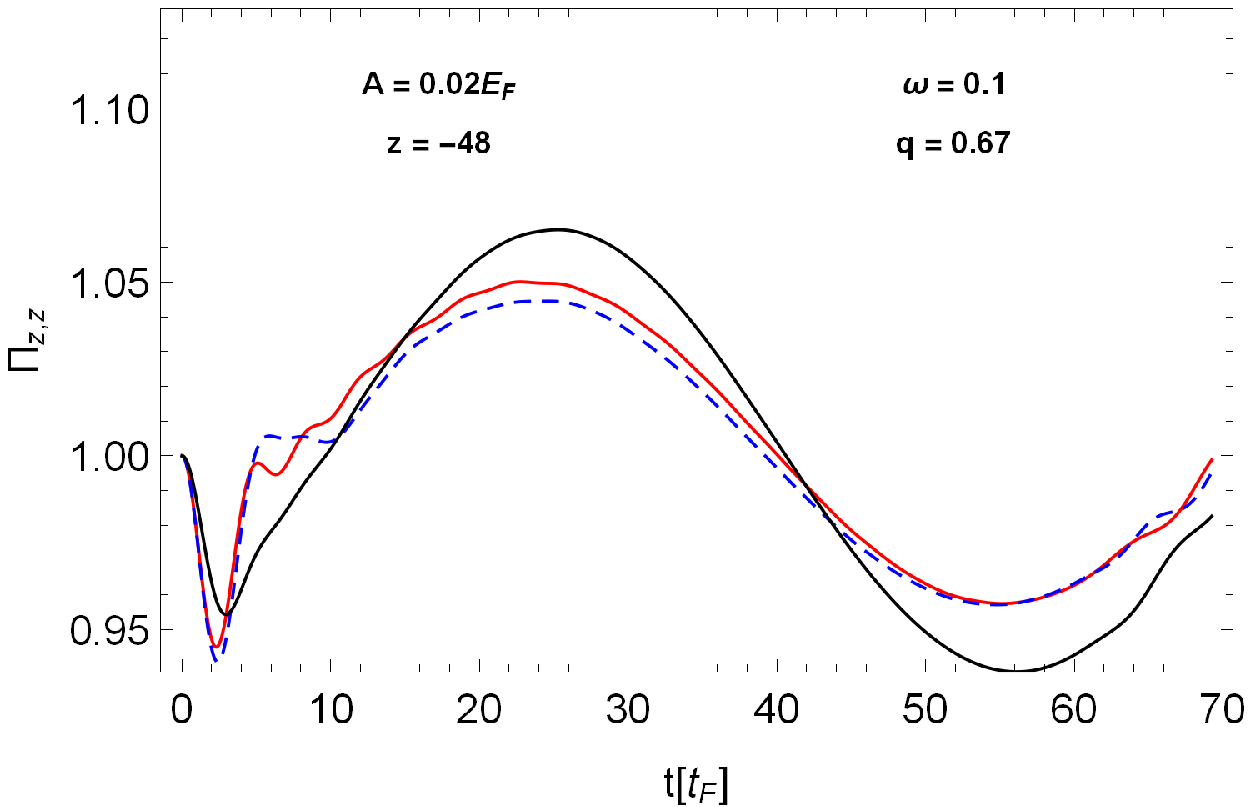}}
	\subfloat{\includegraphics[width = 0.45\linewidth]{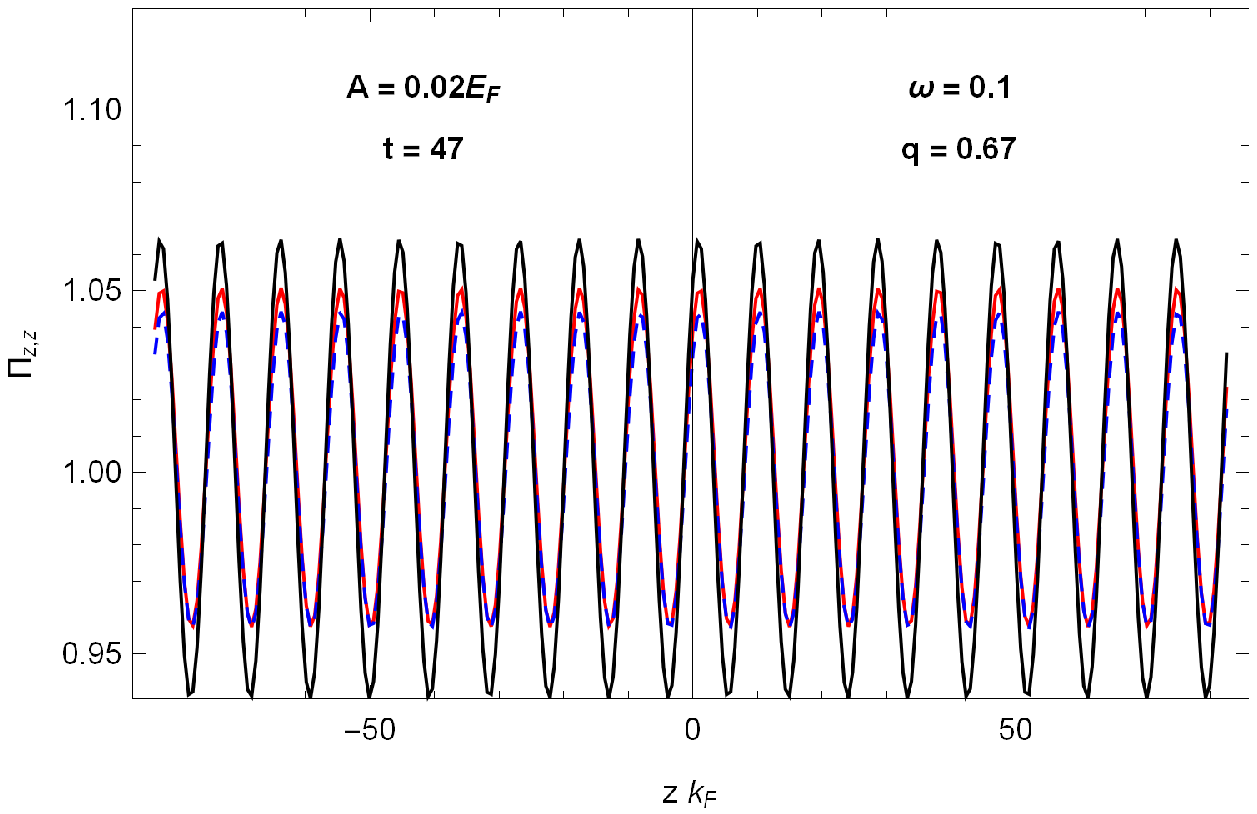}}\\
	\subfloat{\includegraphics[width = 0.45\linewidth]{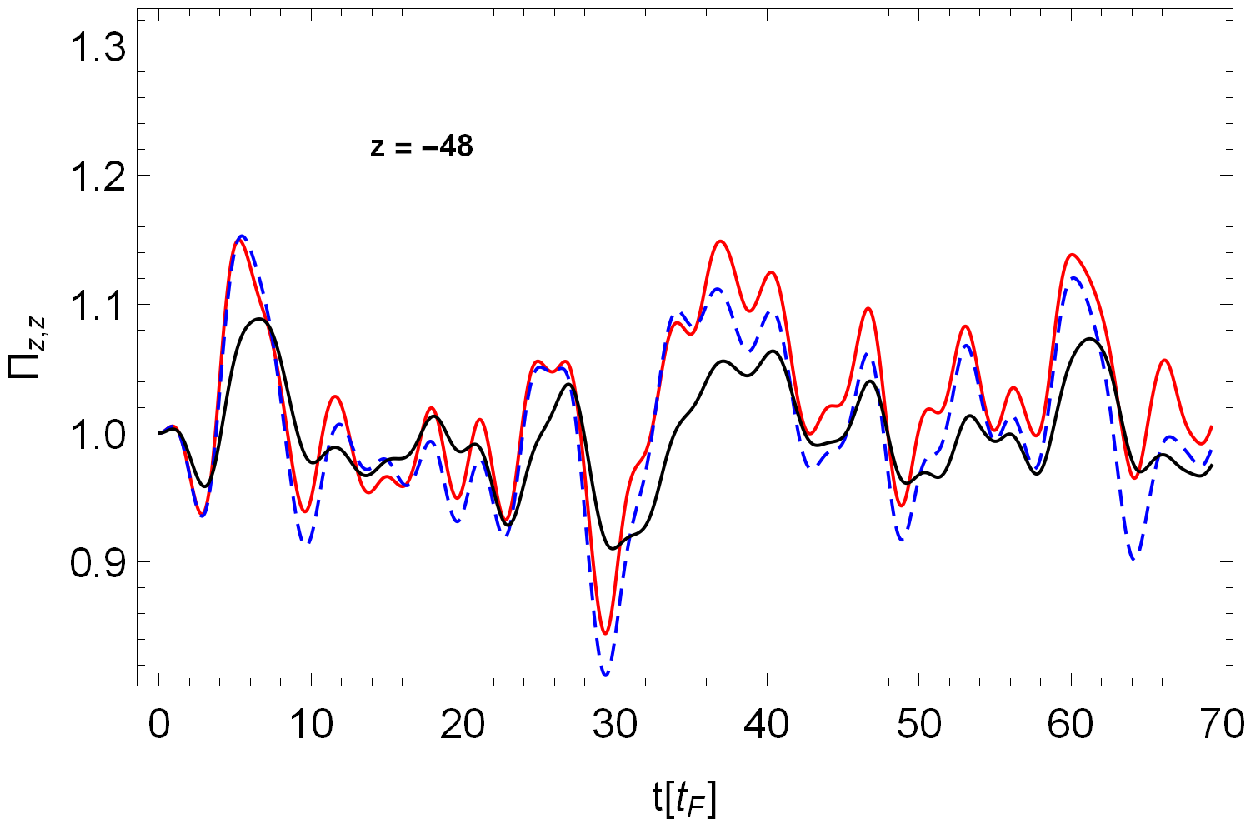}}
    \subfloat{\includegraphics[width = 0.45\linewidth]{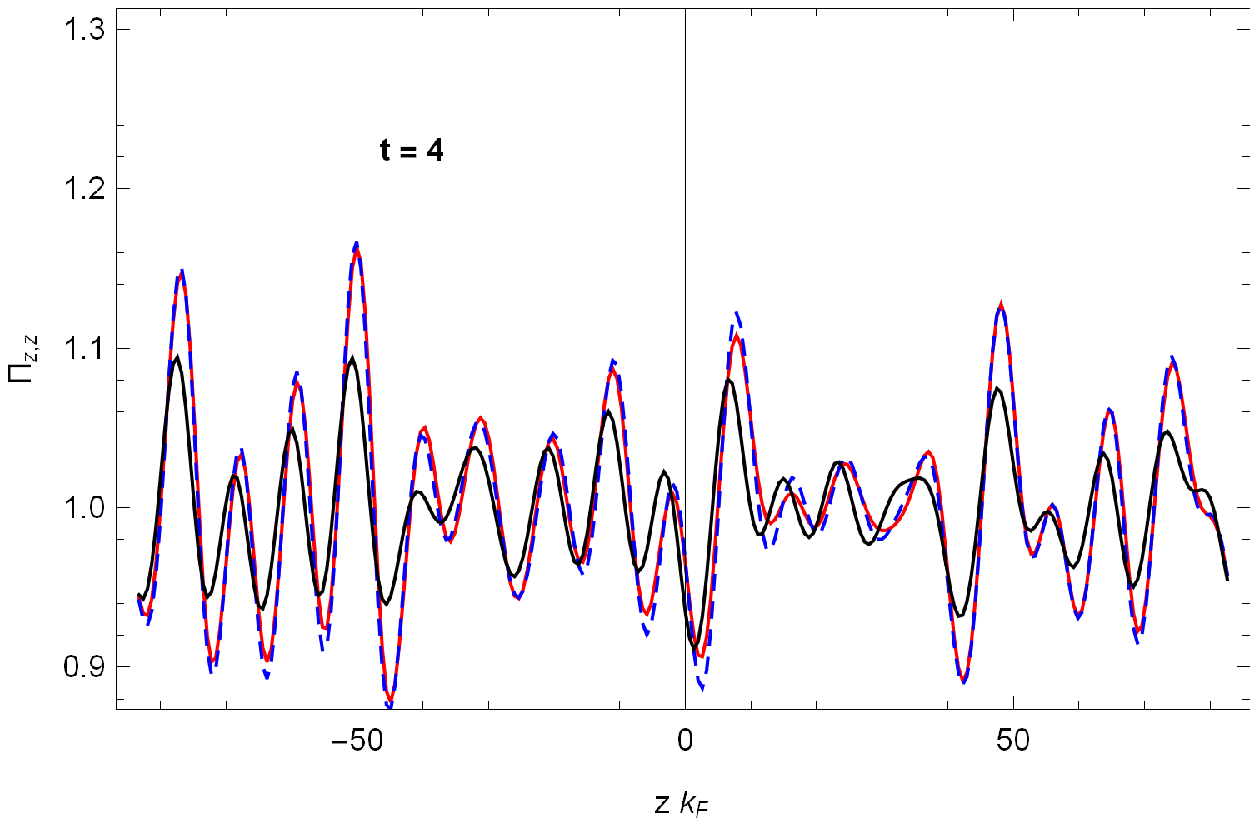}}
	\caption{Temporal (left) and spatial (right) profiles of the pressure at a specific point in space, respectively in time. The profiles are computed via microscopic method (red, full-line), KPDF (blue, dashed) and $\hat{\Pi}_{9/5,1}^0$ (black).}
	\label{Fig_5}
\end{figure}

\emph{The dynamic regime} is the real test for the approximative nature of the KPDF. For that, another ensemble is generated, only now, with frequency dependent modes. In Fig. [\ref{Fig_5}] are plotted the pressure profiles: microscopic (red), the present KPDF (blue, dashed) and $\Pi_{\alpha,1}^0$ (black) for certain spectral modes and at certain time (right) and space (left) points. As expected from the previous analysis, in the asymptotic regions, the results are well reproduced. Moreover, in the intermediate area ($w=1$, $q=0.5$), despite somewhat larger quantitative errors, the qualitative trends are closely followed. In the bottom figure the results of a random potential are presented with the same qualitative/quantitative trends.

\begin{figure}
	\centering
	\includegraphics[width = 0.9\linewidth]{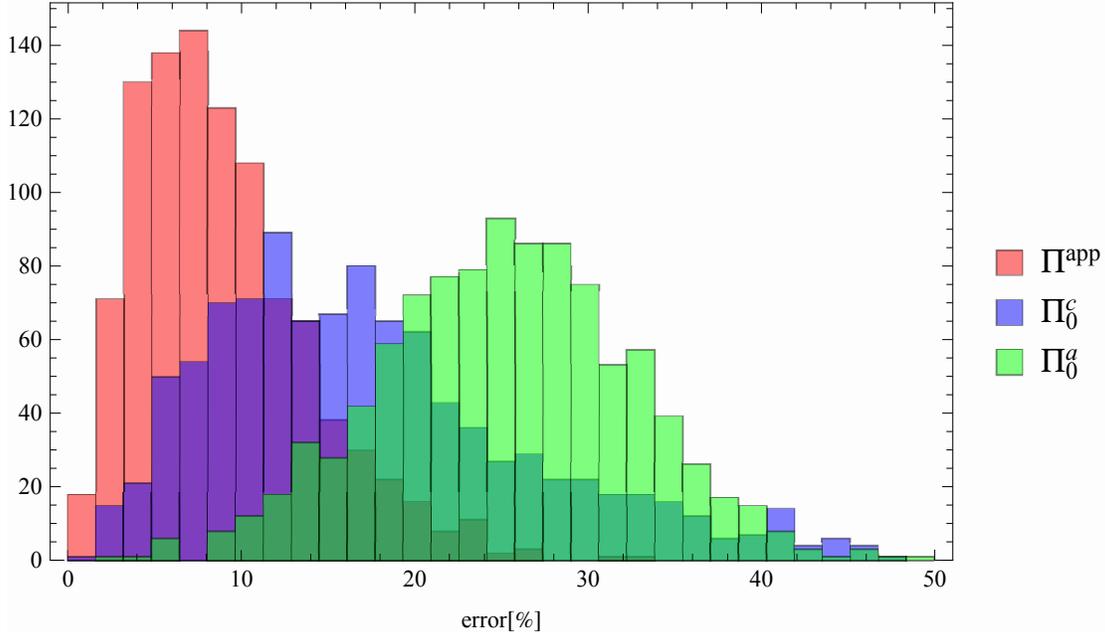}
	\caption{Distribution of errors in the dynamic Monte-Carlo ensemble by KPDF (red), $\hat{\Pi}_0^a$ (green) and $\hat{\Pi}_0^c$ (blue).}
	\label{Fig_6}
\end{figure}

As in the stationary case, we chose to gather the ensemble results in a histogram [\ref{Fig_6}] of the errors averaged over space and time. In contrast with the groundstate, the dynamic regime reveals one order of magnitude larger errors. In the case of the recent \cite{doi:10.1063/1.5003910} $\hat{\Pi}_{9/5,1}^0$, $\hat{\Pi}_{1,1/9}^0$ functionals, this can be understood as a consequence of not being able to deal with certain spectral components of the potential. For the KPDF, errors arise mainly from the particle-hole continuum modes as well as from going beyond the linear regime. Still, it can be concluded at this end that the present KPDF is universally superior to other existing functionals, both for equilibrium and dynamics.

\subsection{Testing on metal clusters}
\label{Sec. IIIc}

In the previous sections, it has been shown that KPDF deals well with a Fermi gas in most spectral cases and slightly above linear dynamics. Still, the agreement is a consequence of working within the thermodynamic limit (large number of particles). There are many systems of interest, especially to nano-plasmonics, which are finite, lack spatial periodicity and contain a small number of electrons.  All these features might hinder the applicability of the KPDF (in particular, the problem of $N$-representability \cite{Levy6062} connected with the low number of fermions). 

To understand what is expected in such scenarios, let us consider as a final test, the case of small spherical sodium clusters $Na_{20}$ and $Na_{40}$. These are conglomerate of sodium atoms, approximatively spherical in which the valence electrons are known to exhibit special behavior. In particular they are known to support electrostatic normal modes relevant for the optical spectra known as localized volume and surface plasmons \cite{RevModPhys.65.677}. The ground state and the normal mode dynamics of these systems is resolved both microscopically (solving the LDA Kohn-Sham equations for the valence electrons within the jellium model) and macroscopically (solving the QHD with the KPDF). Details about the jellium model and numerical implementation of KS Eqns. can be found in \cite{RevModPhys.65.677}. 

\begin{figure}[H]
	\centering
	\subfloat{\includegraphics[width = 0.5\linewidth]{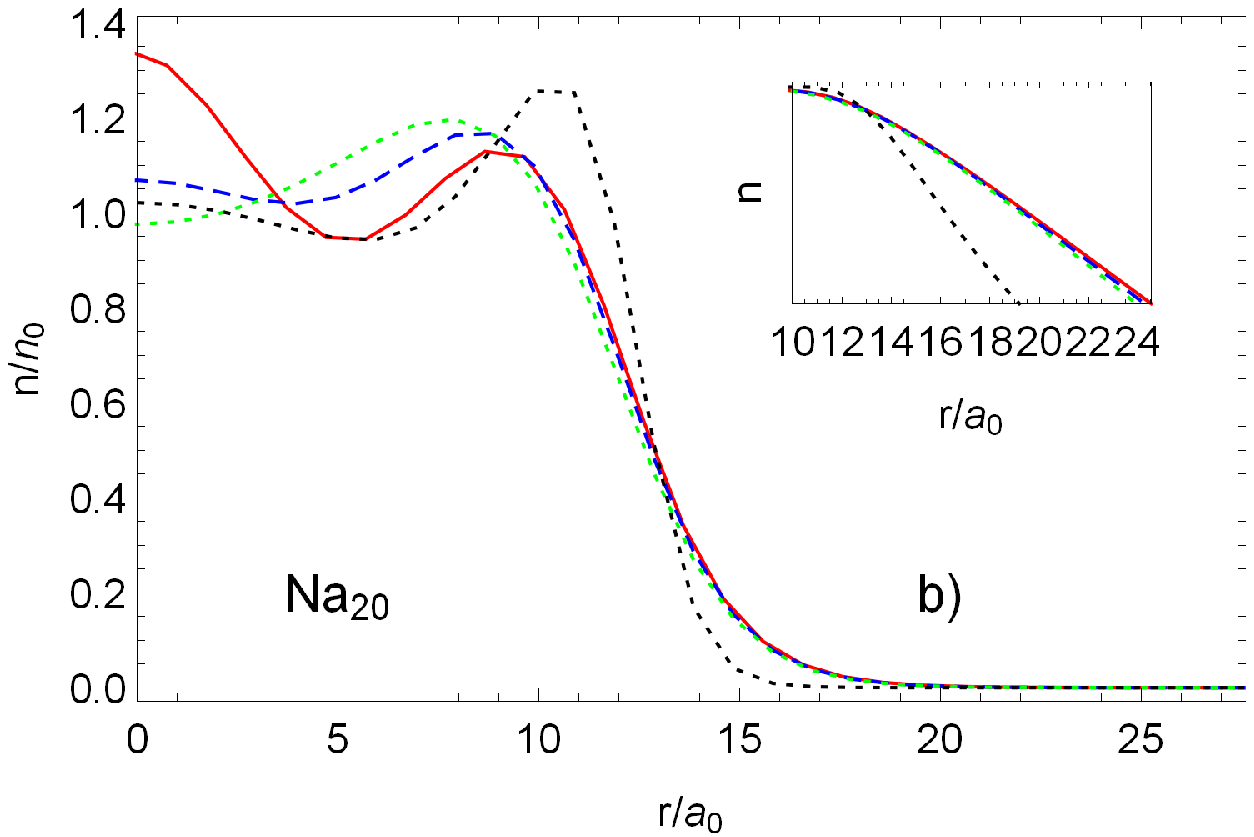}}
	\subfloat{\includegraphics[width = 0.5\linewidth]{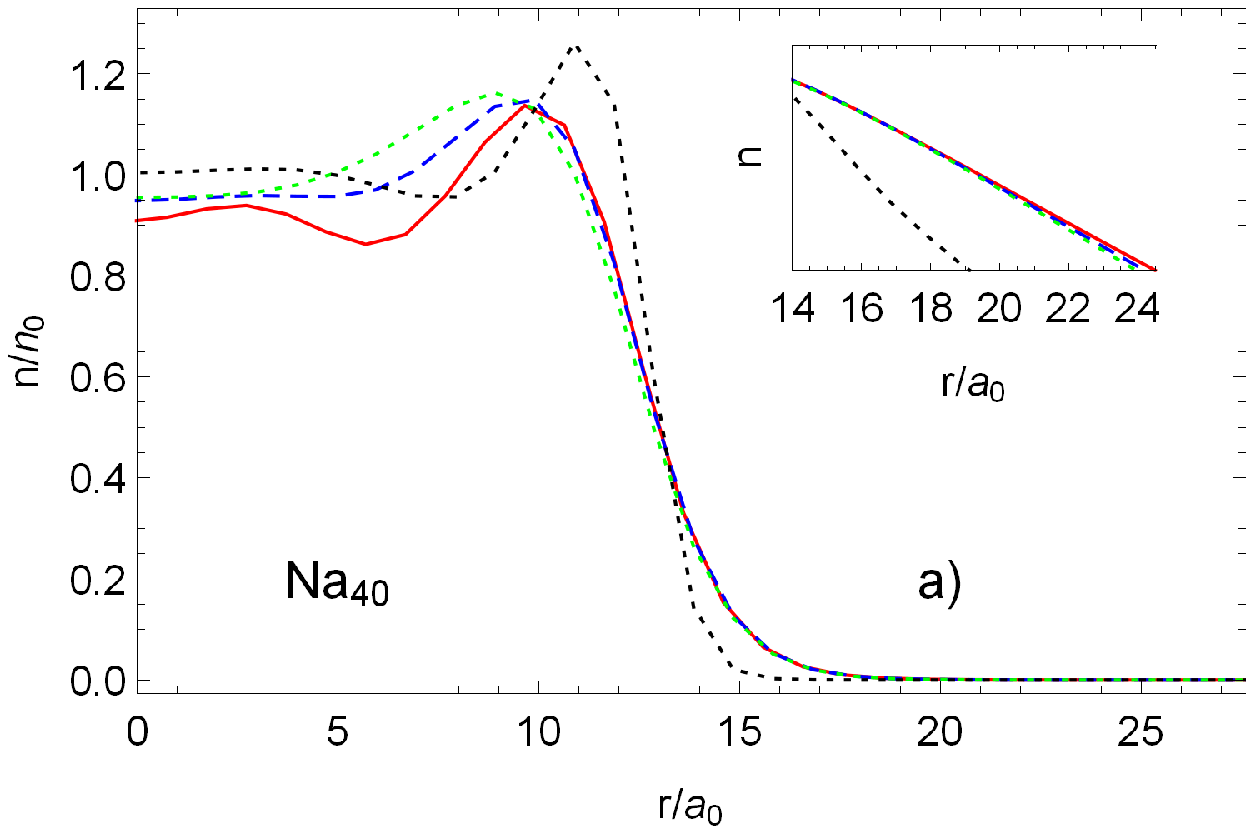}}\\
	\caption{Ground-state electronic density profiles for $Na_{20}$ (left) and $Na_{40}$ (right) computed with: DFT-LDA (red,full-line), KPDF (blue, dashed), $\hat{\Pi}_{1,1/9}^0$ (black, dotted) and $\hat{\Pi}_{1,1}^0$ (green, dotted). Inset: exponential fall of the electronic tail outside the cluster.}
	\label{Fig_7}
\end{figure}

In Fig [\ref{Fig_7}] are plotted the radial profiles of density obtained with KS, KPDF, $\hat{\Pi}_{1,1}^0$ and $\hat{\Pi}_{1,1/9}^0$. As it can be seen, the shell effect in the core of the cluster cannot be reproduced by none of the functionals, this being a direct reflection of the representability problem. This remains a major challenge, for future improvements that can be brought to any functional. Apart from this core behavior, one can see that the KPDF is, again, overall better than other functionals. More important, it is able to capture the correct exponential tail of electronic density outside the cluster (inset of Fig. [\ref{Fig_7}]). This feature is of great importance \cite{PhysRevB.93.205405} in many surface phenomena, such as the static polarizability or the surface plasmon resonance. Regarding the shell oscillations in density, we expect that they get smaller with the size of the cluster, such that, for large clusters, the system is more Fermi gas-like, and the errors should be much smaller.

\begin{figure}[H]
	\centering
	\subfloat{\includegraphics[width = 0.5\linewidth]{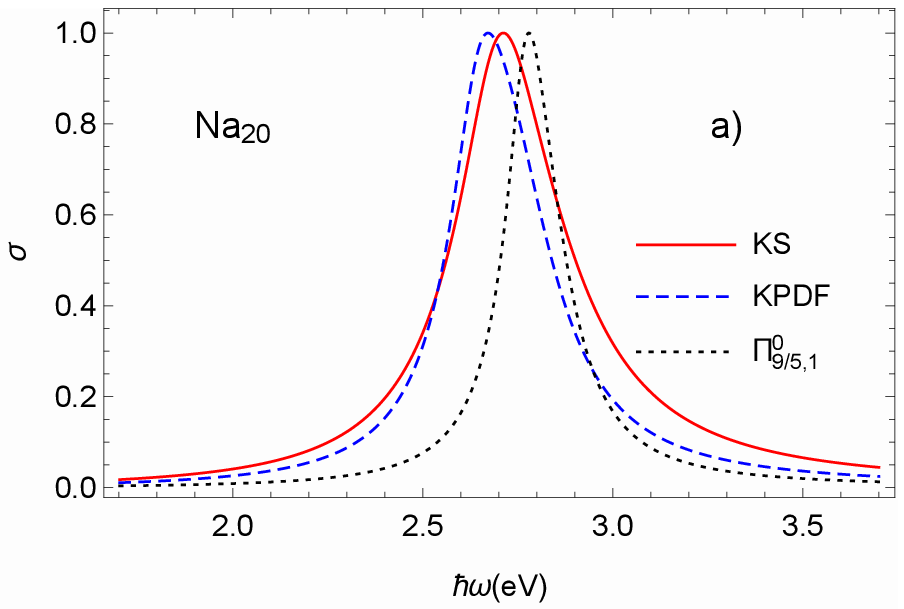}}
	\subfloat{\includegraphics[width = 0.5\linewidth]{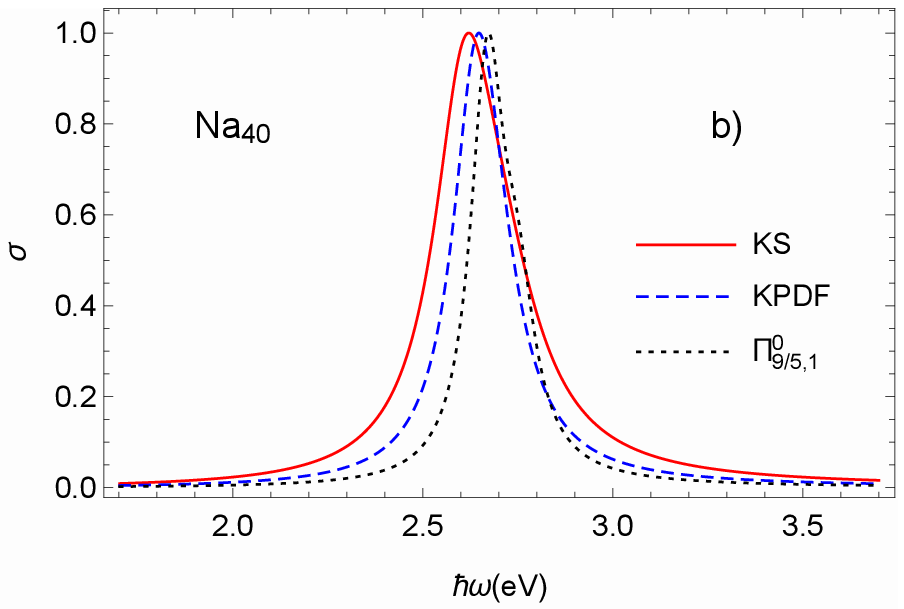}}\\
	\caption{Normalized optical cross-section spectrum for $Na_{20}$ (left) and $Na_{40}$ (right) cluster computed with: KS (DFT-LDA) (red, full-line), KPDF (blue,dashed) and $\hat{\Pi}_{9/5,1}^0$ (black, dotted)}
	\label{Fig_8}
\end{figure}

Finally, the optical spectrum of the clusters is studied in a standard manner: the electrons are excited with a collective initial uniform velocity in the $Oz$ direction. The dynamics under the effect of self-consistent fields is simulated and the dipole moment $d(t)=\int n(\mathbf{r},t)zd\mathbf{r}$ is computed. The optical spectrum is defined as $S(\omega)=\Im d(\omega)$. The results are shown in Fig. [\ref{Fig_8}]. While all functionals give good qualitative agreement with KS-LDA, the KPDF is able to predict the peak of the surface plasmon $10\%$ more accurate than $\hat{\Pi}_{9/5,1}^0$. More important, the width of the peak, i.e. the Landau damping is far better reproduced, given the damping term $\hat{\gamma}$ present in Eq. [\ref{Eq_10}].

\section{Conclusions and discussions}	
\label{Sec. IV}
	
In the present work, a kinetic pressure density functional for a Fermi gas has been derived. The functional is designed to reproduce exactly the homogeneous equilibrium and the linear response accordingly with the kinetic/microscopic theories. While exact in principle, the functional poses a very difficult task due to its space and, more important, time non-locality. To solve this issue, an approximation of the functional has been proposed, reformulating the integral form into a wave-like equation, with the advantage of being local in time.

After spectral analysis, it is shown that the approximation is capable to resolve all asymptotic limits of the frequency-wavelength domain and to interpolate fairly well the intemediate regions, i.e. the particle-hole continuum. In practice, for the Fermi gas, it is able to provide results with one order of magnitude closer to the \emph{exact} microscopic pressure, justifying the numerical effort compared to other existing functionals. Also, at least qualitatively, it is capable to go beyond linear regimes. 

Its main limitation appears as an outcome of the constrains which tacitly assume the thermodynamic limit. Therefore, one of the important features of small, finite systems, the discretness, the shell effect, cannot be reproduced. This problem is connected with a more subtle one, called the $N$-representability problem \cite{Levy6062}. Nonetheless, the results provided by the KPDF are still superior to other functionals, also giving access to a better representation of disipative processes, i.e. Landau damping. 

The purpose of this work goes beyond designing a new functional within the Orbital-Free DFT. It deals with opening a new path to a class of possible functionals. Apart from including other effects as spin, temperature, etc. one could systematically improve the present results in a number of ways. The next logical step would be to constrain the KPDF to reproduce the second order linear response function of a HFG. In this way, the decoupling anzatz used in the derivation would be removed, and also a better congruence with the non-linear regime should be expected. Secondly, one might find simpler or better ways to remove the problem of time non-locality and such, to minimize the errors associated with the particle-hole continuum.


\section{Appendix}	
\label{Sec. V}

\subsection{Pressure decomposition}
\label{Sec. Va}

One can start from the kinetic prescription of the total pressure:
\begin{align*}\hat{\Pi}(\mathbf{r},t)=-\frac{\hbar^2}{4m^2}\lim\limits_{\mathbf{r}'\to \mathbf{r}}(\nabla_\mathbf{r}-\nabla_\mathbf{r'})\otimes(\nabla_\mathbf{r}-\nabla_\mathbf{r'})\rho(\mathbf{r},\mathbf{r}',t)
\end{align*}
And proceed in three steps. First the collective velocity field is isolated by the transformation $\rho(\mathbf{r},\mathbf{r'},t)\equiv e^{iS_0(\mathbf{r},t)}\rho(\mathbf{r},\mathbf{r'},t)e^{-iS_0(\mathbf{r'},t)}$ with $\nabla S_0=m\mathbf{u}(\mathbf{r},t)$ which gives:
\begin{align*}\hat{\Pi}=\frac{\mathbf{j}\otimes\mathbf{j}}{n}-\frac{\hbar^2}{4m^2}\lim\limits_{\mathbf{r}'\to \mathbf{r}}(\nabla_\mathbf{r}-\nabla_\mathbf{r'})\otimes(\nabla_\mathbf{r}-\nabla_\mathbf{r'})\rho(\mathbf{r},\mathbf{r}',t)
\end{align*}
then, together with the identity $$\lim\limits_{\mathbf{r}'\to \mathbf{r}}(\nabla_\mathbf{r}-\nabla_\mathbf{r'})\otimes(\nabla_\mathbf{r}-\nabla_\mathbf{r'})\rho(\mathbf{r},\mathbf{r}',t)=2\nabla\otimes\nabla\rho-\lim\limits_{\mathbf{r}'\to \mathbf{r}}(\nabla_\mathbf{r}\otimes\nabla_\mathbf{r'}+\nabla_\mathbf{r'}\otimes\nabla_\mathbf{r})\rho(\mathbf{r},\mathbf{r}',t)$$  
the transformation $\rho(\mathbf{r},\mathbf{r'},t)\equiv n^{1/2}(\mathbf{r},t)\tilde{\rho}(\mathbf{r},\mathbf{r'},t)n^{1/2}(\mathbf{r'},t)$ is used to get:
\begin{align*}\hat{\Pi}=\frac{\mathbf{j}\otimes\mathbf{j}}{n}+\hat{\Pi}_B[n]+\frac{\hbar^2}{2m^2}n\lim\limits_{\mathbf{r}'\to \mathbf{r}}(\nabla_\mathbf{r}\otimes\nabla_\mathbf{r'}+\nabla_\mathbf{r'}\otimes\nabla_\mathbf{r})\tilde{\rho}(\mathbf{r},\mathbf{r}',t)
\end{align*}
The final transformation is to isolate the TF-like density matrix $\rho_{TF}=3nj_1(k_F|\mathbf{r-r'}|)/(k_F|\mathbf{r-r'}|)$ from $\tilde{\rho}$ to get:
\begin{align*}&\hat{\Pi}=\frac{\mathbf{j}\otimes\mathbf{j}}{n}+\hat{\Pi}_B[n]+\hat{\Pi}_{TF}[n]+\hat{\Pi}_{NL}[n]\\
&\hat{\Pi}_{NL}[n]=\frac{\hbar^2}{2m^2}n\lim\limits_{\mathbf{r}'\to \mathbf{r}}(\nabla_\mathbf{r}\otimes\nabla_\mathbf{r'}+\nabla_\mathbf{r'}\otimes\nabla_\mathbf{r})\tilde{\tilde{\rho}}(\mathbf{r},\mathbf{r}',t)
\end{align*}
The non-local pressure is rearranged as in integral form over space-time with non-local density dependent kernel:
\begin{align*}
	&\hat{\Pi}_{NL}=\frac{\hbar^2}{2m^2}\int dx'\delta(x-x')(\nabla_{\mathbf{r}}\otimes\nabla_{\mathbf{r}'}+\nabla_{\mathbf{r'}}\otimes\nabla_{\mathbf{r}})\tilde{\rho}(x,x')\\
	&\tilde{\rho}(x,x')=\int dy dy'n^{1/2}(y)\mathcal{O}(x,y,x',y';n(x),n(x'))n^{1/2}(y)
\end{align*}
where $x=(\mathbf{r},t)$, $y=(\mathbf{r'},t')$.

While a supplementary dependency on density could be imposed as $\mathcal{O}[x,y,x',y';n(x),n(x'), n(y),n(y')]$, in practice, this would only make the functional harder to implement without any improvements. The time-space invariance of $n_0$ implies $\mathcal{O}(x,y,x',y;n_0,n_0)\equiv\mathcal{O}(x-y,x'-y';n_0,n_0)$ therefore a Fourier representation ($\xi=(\omega, k)$) of the kernel is chosen in order to apply the linear response condition:
$$\mathcal{O}(x-y,x'-y';n_1,n_2)=\int d\xi d\zeta e^{-i\xi(x-y)}\mathcal{\tilde{O}}(\xi,\zeta;n_1,n_2)e^{-i\zeta(x'-y')}$$
The constrains \ref{Cond1},\ref{Cond2} can be reformulated for the non-local pressure as:
\begin{align*}
\Pi_{NL}[n_0]=0\\
\mathfrak{F}\{\left(\frac{\delta \Pi_{NL}}{\delta n}\right)_{n_0}\}=\phi(\xi)=\frac{\omega^2}{k^2}-\frac{n_0}{m\chi(\omega,k)}-\frac{\delta \Pi_B}{\delta n}-\frac{\delta \Pi_{TF}}{\delta n}
\end{align*}
which in terms of kernel $\tilde{\mathcal{O}}$ can be rewritten after some calculus as:
\begin{align*}
&\lim\limits_{\substack{\zeta_1\to 0\\ \zeta_2\to 0}}(\underline{\zeta_1}\otimes\underline{\zeta_2}+\underline{\zeta_2}\otimes\underline{\zeta_1})\mathcal{\tilde{O}}(\zeta_1,\zeta_2;n_0,n_0)=0\nonumber\\
\frac{\hbar^2}{m^2}&\lim\limits_{\substack{\zeta_1\to \xi\\ \zeta_2\to 0}}(\underline{\zeta_1}\otimes\underline{\zeta_2}+\underline{\zeta_2}\otimes\underline{\zeta_1})\mathcal{\tilde{O}}(\zeta_1,\zeta_2;n_0,n_0)=\phi(\xi)
\end{align*}
where $\underline{\xi}=\mathbf{k}$, the spatial components of the quadri-vector $\xi$. A supplementary anzatz is used to remove the null behavior at the origin and decouple the integrals:
\begin{equation*}\label{anzatz}
\mathcal{O}(\xi,\zeta;n_1,n_2)\equiv\{\mathcal{O}(\xi;n_1)+\mathcal{O}(\zeta;n_2)\}/(\underline{\xi}\cdot\underline{\zeta})
\end{equation*}
using these in the formula for $\Pi_{NL}$ one gets:
\begin{align*}
&\hat{\Pi}_{NL}[n]=\int dx'\mathcal{L}(x,x')[n^{1/2}(x)\mathcal{D}(x')+n^{1/2}(x')\mathcal{D}(x)]\nonumber\\
&\mathcal{L}(x,x')=\delta(x-x')(\nabla_{\mathbf{r}}\otimes\nabla_{\mathbf{r}'}+\nabla_{\mathbf{r'}}\otimes\nabla_{\mathbf{r}})(\nabla_{\mathbf{r}}\nabla_{\mathbf{r'}})^{-1}\nonumber\\
&\mathcal{D}(x)=\frac{1}{2}\int dy \int d\xi e^{-i\xi(x-y)}\phi(\xi;n(x))n^{1/2}(y)
\end{align*}

\subsection{Approximating the kernel}
\label{Sec. Vb}

The Drude model for the dielectric function is well known:
$$\varepsilon=1-\frac{\omega_p}{\omega^2+i\omega\gamma}$$
Since the dielectric function is related to the LRF, and implicitely, to the kernel $\phi$ by $\varepsilon=1-v(k)\chi(\omega,k)$, the following approximative form is proposed:
\begin{align*}
\mathcal{\phi}^{app}(w,q)\approx \phi_0^\infty\frac{w^2-i\gamma(q) w+t_2(q)}{w^2-i\gamma(q)w+t_1(q)}\\
\end{align*}
where the scaled spectral variables $w=\hbar^2k_F^2/(2m\omega)$, $q=k/k_F$ have been used. This approximation is design to express the $w$ dependency as a rational function and to reproduce \emph{simultaneously} the asymptotic limits $\omega=0$, $\omega\to\infty$ as well as the mid-line within the particle-hole continuum $w=q^2+q$. 
Rewriting the convolution expression for $\mathcal{D}$ as: $\mathcal{D}=\phi^{app}\otimes n^{1/2}$, applying the Fourier transform, rearrenging the terms and applying an inverse Fourier transform, one can write down an wave equation for $\mathcal{D}$:
\begin{eqnarray}
[\partial_{t,t}-\hat{\gamma}\partial_t-\hat{t}_1]\mathcal{D}=\frac{4}{15}k_F^2 [\partial_{t,t}-\hat{\gamma}\partial_t-\hat{t}_2]n^{1/2}
\end{eqnarray}
where, through the LDA limit for 3D systems $k_F=(3\pi^2n)^{1/3}$ and the spatial operators $\gamma, t_1,t_2$ are applied accordingly with the above Fourier prescription as:

$$\hat{\gamma}f(\mathbf{r})=\int d\mathbf{r'}\left(\int d\mathbf{k}e^{i\mathbf{k}(\mathbf{r}-\mathbf{r}')}\gamma(\frac{\mathbf{k}}{k_F(\mathbf{r})})\right)f(\mathbf{r'})$$
The functions $\gamma, t_1,t_2, \phi_0^0,\phi_1^\infty$ are defined analytically through the hierarchy:
\begin{align*}
&\Psi_3(z)=\frac{1}{4} \left(1-z^2\right) \log \left(\frac{z+1}{z-1}\right)+\frac{z}{2}\\
&\tilde{\chi}(w,q)=\frac{3 \left(\Psi _3\left(\frac{w}{2 q}-\frac{q}{2}\right)-\Psi _3\left(\frac{w}{2 q}+\frac{q}{2}\right)\right)}{q}\\
&\phi(w,q)=\frac{w^2}{4 q^2}-\frac{q^2}{4}-\frac{1}{3}-\frac{1}{\tilde{\chi}(w,q)}\\
&\lim\limits_{w\to\infty}\phi(w,z)=\phi_0^\infty+\phi_1^\infty(q)/w^2\hspace{0.3cm};\hspace{0.3cm}
\phi_1^\infty(q)=\frac{4 q^4}{5}+\frac{48 q^2}{175}\hspace{0.3cm};\hspace{0.3cm}\phi_0^\infty=\frac{4}{15}\\
&\phi_0^0(q)=\phi(0,q)\hspace{0.3cm};\hspace{0.3cm}t_1(q)=\frac{\phi_1^\infty(q)}{\phi_0^0(q)-\phi_0^\infty}\\
&t_2(q)=t_1(q)\frac{\phi_0^0(q)}{\phi_0^\infty}\hspace{0.3cm};\hspace{0.3cm}\phi_0^0(q)\equiv \phi(\omega=0,q)\\
&\gamma(q)=\frac{-i}{q^2+q}\left((q^2+q)^2+\phi_1^\infty(q)\frac{\phi_0^0(q)-\phi(q^2+q,q)}{(\phi_0^0(q)-\phi_0^\infty)(\phi(q^2+q,q)-\phi_0^\infty)}\right)\end{align*}

We note that, in the limit of small spatial oscillations $|\nabla \ln n|\ll k_F$:
\begin{align*}
&\hat{\gamma}f(\mathbf{r})\approx \frac{0.87-0.5i}{k_F}|\nabla|f(\mathbf{r})+\frac{0.28+0.5i}{k_F^2}\nabla^2f(\mathbf{r})\\
&\hat{t}_2f(\mathbf{r})\approx \frac{6}{7k_F^4}\nabla^4f(\mathbf{r})\\
&\hat{t}_1f(\mathbf{r})\approx \frac{36}{35k_F^2}\nabla^2f(\mathbf{r})-\frac{15}{7k_F^4}\nabla^4f(\mathbf{r})
\end{align*}

\bibliographystyle{aipsamp}
\bibliography{biblio}

\begin{thebibliography}{43}
\expandafter\ifx\csname natexlab\endcsname\relax\def\natexlab#1{#1}\fi
\expandafter\ifx\csname bibnamefont\endcsname\relax
  \def\bibnamefont#1{#1}\fi
\expandafter\ifx\csname bibfnamefont\endcsname\relax
  \def\bibfnamefont#1{#1}\fi
\expandafter\ifx\csname citenamefont\endcsname\relax
  \def\citenamefont#1{#1}\fi
\expandafter\ifx\csname url\endcsname\relax
  \def\url#1{\texttt{#1}}\fi
\expandafter\ifx\csname urlprefix\endcsname\relax\def\urlprefix{URL }\fi
\providecommand{\bibinfo}[2]{#2}
\providecommand{\eprint}[2][]{\url{#2}}

\bibitem[{\citenamefont{Fletcher et~al.}(2015)\citenamefont{Fletcher, Lee,
  D{\"o}ppner, Galtier, Nagler, Heimann, Fortmann, LePape, Ma, Millot
  et~al.}}]{Fletcher2015}
\bibinfo{author}{\bibfnamefont{L.~B.} \bibnamefont{Fletcher}},
  \bibinfo{author}{\bibfnamefont{H.~J.} \bibnamefont{Lee}},
  \bibinfo{author}{\bibfnamefont{T.}~\bibnamefont{D{\"o}ppner}},
  \bibinfo{author}{\bibfnamefont{E.}~\bibnamefont{Galtier}},
  \bibinfo{author}{\bibfnamefont{B.}~\bibnamefont{Nagler}},
  \bibinfo{author}{\bibfnamefont{P.}~\bibnamefont{Heimann}},
  \bibinfo{author}{\bibfnamefont{C.}~\bibnamefont{Fortmann}},
  \bibinfo{author}{\bibfnamefont{S.}~\bibnamefont{LePape}},
  \bibinfo{author}{\bibfnamefont{T.}~\bibnamefont{Ma}},
  \bibinfo{author}{\bibfnamefont{M.}~\bibnamefont{Millot}},
  \bibnamefont{et~al.}, \bibinfo{journal}{Nature Photonics}
  \textbf{\bibinfo{volume}{9}}, \bibinfo{pages}{274 EP }
  (\bibinfo{year}{2015}), \bibinfo{note}{article},
  \urlprefix\url{http://dx.doi.org/10.1038/nphoton.2015.41}.

\bibitem[{\citenamefont{Scholl et~al.}(2012)\citenamefont{Scholl, Koh, and
  Dionne}}]{Scholl2012}
\bibinfo{author}{\bibfnamefont{J.~A.} \bibnamefont{Scholl}},
  \bibinfo{author}{\bibfnamefont{A.~L.} \bibnamefont{Koh}}, \bibnamefont{and}
  \bibinfo{author}{\bibfnamefont{J.~A.} \bibnamefont{Dionne}},
  \bibinfo{journal}{Nature} \textbf{\bibinfo{volume}{483}}, \bibinfo{pages}{421
  EP } (\bibinfo{year}{2012}), \bibinfo{note}{article},
  \urlprefix\url{http://dx.doi.org/10.1038/nature10904}.

\bibitem[{\citenamefont{Calvayrac et~al.}(2000)\citenamefont{Calvayrac,
  Reinhard, Suraud, and Ullrich}}]{CALVAYRAC2000493}
\bibinfo{author}{\bibfnamefont{F.}~\bibnamefont{Calvayrac}},
  \bibinfo{author}{\bibfnamefont{P.-G.} \bibnamefont{Reinhard}},
  \bibinfo{author}{\bibfnamefont{E.}~\bibnamefont{Suraud}}, \bibnamefont{and}
  \bibinfo{author}{\bibfnamefont{C.}~\bibnamefont{Ullrich}},
  \bibinfo{journal}{Physics Reports} \textbf{\bibinfo{volume}{337}},
  \bibinfo{pages}{493 } (\bibinfo{year}{2000}), ISSN \bibinfo{issn}{0370-1573},
  \urlprefix\url{http://www.sciencedirect.com/science/article/pii/S0370157300000430}.

\bibitem[{\citenamefont{Crouseilles et~al.}(2008)\citenamefont{Crouseilles,
  Hervieux, and Manfredi}}]{PhysRevB.78.155412}
\bibinfo{author}{\bibfnamefont{N.}~\bibnamefont{Crouseilles}},
  \bibinfo{author}{\bibfnamefont{P.-A.} \bibnamefont{Hervieux}},
  \bibnamefont{and} \bibinfo{author}{\bibfnamefont{G.}~\bibnamefont{Manfredi}},
  \bibinfo{journal}{Phys. Rev. B} \textbf{\bibinfo{volume}{78}},
  \bibinfo{pages}{155412} (\bibinfo{year}{2008}),
  \urlprefix\url{https://link.aps.org/doi/10.1103/PhysRevB.78.155412}.

\bibitem[{\citenamefont{Fortney}()}]{doi:10.1002/ctpp.201300001}
\bibinfo{author}{\bibfnamefont{J.~J.} \bibnamefont{Fortney}},
  \bibinfo{journal}{Contributions to Plasma Physics}
  \textbf{\bibinfo{volume}{53}}, \bibinfo{pages}{385} (????),
  \eprint{https://onlinelibrary.wiley.com/doi/pdf/10.1002/ctpp.201300001},
  \urlprefix\url{https://onlinelibrary.wiley.com/doi/abs/10.1002/ctpp.201300001}.

\bibitem[{\citenamefont{Tame et~al.}(2013)\citenamefont{Tame, McEnery,
  {\"O}zdemir, Lee, Maier, and Kim}}]{Tame2013}
\bibinfo{author}{\bibfnamefont{M.~S.} \bibnamefont{Tame}},
  \bibinfo{author}{\bibfnamefont{K.~R.} \bibnamefont{McEnery}},
  \bibinfo{author}{\bibfnamefont{S.~K.} \bibnamefont{{\"O}zdemir}},
  \bibinfo{author}{\bibfnamefont{J.}~\bibnamefont{Lee}},
  \bibinfo{author}{\bibfnamefont{S.~A.} \bibnamefont{Maier}}, \bibnamefont{and}
  \bibinfo{author}{\bibfnamefont{M.~S.} \bibnamefont{Kim}},
  \bibinfo{journal}{Nature Physics} \textbf{\bibinfo{volume}{9}},
  \bibinfo{pages}{329 EP } (\bibinfo{year}{2013}), \bibinfo{note}{review
  Article}, \urlprefix\url{http://dx.doi.org/10.1038/nphys2615}.

\bibitem[{\citenamefont{Schuller et~al.}(2010)\citenamefont{Schuller, Barnard,
  Cai, Jun, White, and Brongersma}}]{Schuller2010}
\bibinfo{author}{\bibfnamefont{J.~A.} \bibnamefont{Schuller}},
  \bibinfo{author}{\bibfnamefont{E.~S.} \bibnamefont{Barnard}},
  \bibinfo{author}{\bibfnamefont{W.}~\bibnamefont{Cai}},
  \bibinfo{author}{\bibfnamefont{Y.~C.} \bibnamefont{Jun}},
  \bibinfo{author}{\bibfnamefont{J.~S.} \bibnamefont{White}}, \bibnamefont{and}
  \bibinfo{author}{\bibfnamefont{M.~L.} \bibnamefont{Brongersma}},
  \bibinfo{journal}{Nature Materials} \textbf{\bibinfo{volume}{9}},
  \bibinfo{pages}{193 EP } (\bibinfo{year}{2010}), \bibinfo{note}{review
  Article}, \urlprefix\url{http://dx.doi.org/10.1038/nmat2630}.

\bibitem[{\citenamefont{Gramotnev and Bozhevolnyi}(2010)}]{Gramotnev2010}
\bibinfo{author}{\bibfnamefont{D.~K.} \bibnamefont{Gramotnev}}
  \bibnamefont{and} \bibinfo{author}{\bibfnamefont{S.~I.}
  \bibnamefont{Bozhevolnyi}}, \bibinfo{journal}{Nature Photonics}
  \textbf{\bibinfo{volume}{4}}, \bibinfo{pages}{83 EP } (\bibinfo{year}{2010}),
  \bibinfo{note}{review Article},
  \urlprefix\url{http://dx.doi.org/10.1038/nphoton.2009.282}.

\bibitem[{\citenamefont{Tsintsadze and
  Tsintsadze}(2009)}]{0295-5075-88-3-35001}
\bibinfo{author}{\bibfnamefont{N.~L.} \bibnamefont{Tsintsadze}}
  \bibnamefont{and} \bibinfo{author}{\bibfnamefont{L.~N.}
  \bibnamefont{Tsintsadze}}, \bibinfo{journal}{EPL (Europhysics Letters)}
  \textbf{\bibinfo{volume}{88}}, \bibinfo{pages}{35001} (\bibinfo{year}{2009}),
  \urlprefix\url{http://stacks.iop.org/0295-5075/88/i=3/a=35001}.

\bibitem[{\citenamefont{Haas et~al.}(2010)\citenamefont{Haas, Marklund, Brodin,
  and Zamanian}}]{HAAS2010481}
\bibinfo{author}{\bibfnamefont{F.}~\bibnamefont{Haas}},
  \bibinfo{author}{\bibfnamefont{M.}~\bibnamefont{Marklund}},
  \bibinfo{author}{\bibfnamefont{G.}~\bibnamefont{Brodin}}, \bibnamefont{and}
  \bibinfo{author}{\bibfnamefont{J.}~\bibnamefont{Zamanian}},
  \bibinfo{journal}{Physics Letters A} \textbf{\bibinfo{volume}{374}},
  \bibinfo{pages}{481 } (\bibinfo{year}{2010}), ISSN \bibinfo{issn}{0375-9601},
  \urlprefix\url{http://www.sciencedirect.com/science/article/pii/S0375960109014248}.

\bibitem[{\citenamefont{McWeeny}(1992)}]{9780124865525}
\bibinfo{author}{\bibfnamefont{R.}~\bibnamefont{McWeeny}},
  \emph{\bibinfo{title}{Methods of Molecular Quantum Mechanics}}
  (\bibinfo{publisher}{Academic Press}, \bibinfo{year}{1992}), ISBN
  \bibinfo{isbn}{0124865526},
  \urlprefix\url{https://www.amazon.com/Methods-Molecular-Quantum-Mechanics-McWeeny/dp/0124865526?SubscriptionId=AKIAIOBINVZYXZQZ2U3A&tag=chimbori05-20&linkCode=xm2&camp=2025&creative=165953&creativeASIN=0124865526}.

\bibitem[{\citenamefont{Runge and Gross}(1984)}]{PhysRevLett.52.997}
\bibinfo{author}{\bibfnamefont{E.}~\bibnamefont{Runge}} \bibnamefont{and}
  \bibinfo{author}{\bibfnamefont{E.~K.~U.} \bibnamefont{Gross}},
  \bibinfo{journal}{Phys. Rev. Lett.} \textbf{\bibinfo{volume}{52}},
  \bibinfo{pages}{997} (\bibinfo{year}{1984}),
  \urlprefix\url{https://link.aps.org/doi/10.1103/PhysRevLett.52.997}.

\bibitem[{\citenamefont{Gardner}(1994)}]{doi:10.1137/S0036139992240425}
\bibinfo{author}{\bibfnamefont{C.}~\bibnamefont{Gardner}},
  \bibinfo{journal}{SIAM Journal on Applied Mathematics}
  \textbf{\bibinfo{volume}{54}}, \bibinfo{pages}{409} (\bibinfo{year}{1994}),
  \eprint{https://doi.org/10.1137/S0036139992240425},
  \urlprefix\url{https://doi.org/10.1137/S0036139992240425}.

\bibitem[{\citenamefont{Domps et~al.}(1998)\citenamefont{Domps, Reinhard, and
  Suraud}}]{PhysRevLett.80.5520}
\bibinfo{author}{\bibfnamefont{A.}~\bibnamefont{Domps}},
  \bibinfo{author}{\bibfnamefont{P.-G.} \bibnamefont{Reinhard}},
  \bibnamefont{and} \bibinfo{author}{\bibfnamefont{E.}~\bibnamefont{Suraud}},
  \bibinfo{journal}{Phys. Rev. Lett.} \textbf{\bibinfo{volume}{80}},
  \bibinfo{pages}{5520} (\bibinfo{year}{1998}),
  \urlprefix\url{https://link.aps.org/doi/10.1103/PhysRevLett.80.5520}.

\bibitem[{\citenamefont{Michta et~al.}()\citenamefont{Michta, Graziani, and
  Bonitz}}]{doi:10.1002/ctpp.201500024}
\bibinfo{author}{\bibfnamefont{D.}~\bibnamefont{Michta}},
  \bibinfo{author}{\bibfnamefont{F.}~\bibnamefont{Graziani}}, \bibnamefont{and}
  \bibinfo{author}{\bibfnamefont{M.}~\bibnamefont{Bonitz}},
  \bibinfo{journal}{Contributions to Plasma Physics}
  \textbf{\bibinfo{volume}{55}}, \bibinfo{pages}{437} (????),
  \eprint{https://onlinelibrary.wiley.com/doi/pdf/10.1002/ctpp.201500024},
  \urlprefix\url{https://onlinelibrary.wiley.com/doi/abs/10.1002/ctpp.201500024}.

\bibitem[{\citenamefont{Palade and Baran}(2015)}]{0953-4075-48-18-185102}
\bibinfo{author}{\bibfnamefont{D.~I.} \bibnamefont{Palade}} \bibnamefont{and}
  \bibinfo{author}{\bibfnamefont{V.}~\bibnamefont{Baran}},
  \bibinfo{journal}{Journal of Physics B: Atomic, Molecular and Optical
  Physics} \textbf{\bibinfo{volume}{48}}, \bibinfo{pages}{185102}
  (\bibinfo{year}{2015}),
  \urlprefix\url{http://stacks.iop.org/0953-4075/48/i=18/a=185102}.

\bibitem[{\citenamefont{Moldabekov et~al.}(2018)\citenamefont{Moldabekov,
  Bonitz, and Ramazanov}}]{doi:10.1063/1.5003910}
\bibinfo{author}{\bibfnamefont{Z.~A.} \bibnamefont{Moldabekov}},
  \bibinfo{author}{\bibfnamefont{M.}~\bibnamefont{Bonitz}}, \bibnamefont{and}
  \bibinfo{author}{\bibfnamefont{T.~S.} \bibnamefont{Ramazanov}},
  \bibinfo{journal}{Physics of Plasmas} \textbf{\bibinfo{volume}{25}},
  \bibinfo{pages}{031903} (\bibinfo{year}{2018}),
  \eprint{https://doi.org/10.1063/1.5003910},
  \urlprefix\url{https://doi.org/10.1063/1.5003910}.

\bibitem[{\citenamefont{Manfredi and Haas}(2001)}]{PhysRevB.64.075316}
\bibinfo{author}{\bibfnamefont{G.}~\bibnamefont{Manfredi}} \bibnamefont{and}
  \bibinfo{author}{\bibfnamefont{F.}~\bibnamefont{Haas}},
  \bibinfo{journal}{Phys. Rev. B} \textbf{\bibinfo{volume}{64}},
  \bibinfo{pages}{075316} (\bibinfo{year}{2001}),
  \urlprefix\url{https://link.aps.org/doi/10.1103/PhysRevB.64.075316}.

\bibitem[{\citenamefont{Myers and Swiatecki}(1996)}]{MYERS1996141}
\bibinfo{author}{\bibfnamefont{W.}~\bibnamefont{Myers}} \bibnamefont{and}
  \bibinfo{author}{\bibfnamefont{W.}~\bibnamefont{Swiatecki}},
  \bibinfo{journal}{Nuclear Physics A} \textbf{\bibinfo{volume}{601}},
  \bibinfo{pages}{141 } (\bibinfo{year}{1996}), ISSN \bibinfo{issn}{0375-9474},
  \urlprefix\url{http://www.sciencedirect.com/science/article/pii/0375947495005099}.

\bibitem[{\citenamefont{Chu et~al.}(1977)\citenamefont{Chu, Jennings, and
  Brack}}]{CHU1977407}
\bibinfo{author}{\bibfnamefont{Y.}~\bibnamefont{Chu}},
  \bibinfo{author}{\bibfnamefont{B.}~\bibnamefont{Jennings}}, \bibnamefont{and}
  \bibinfo{author}{\bibfnamefont{M.}~\bibnamefont{Brack}},
  \bibinfo{journal}{Physics Letters B} \textbf{\bibinfo{volume}{68}},
  \bibinfo{pages}{407 } (\bibinfo{year}{1977}), ISSN \bibinfo{issn}{0370-2693},
  \urlprefix\url{http://www.sciencedirect.com/science/article/pii/0370269377904543}.

\bibitem[{\citenamefont{Clerouin et~al.}(1992)\citenamefont{Clerouin, Pollock,
  and Zerah}}]{PhysRevA.46.5130}
\bibinfo{author}{\bibfnamefont{J.}~\bibnamefont{Clerouin}},
  \bibinfo{author}{\bibfnamefont{E.~L.} \bibnamefont{Pollock}},
  \bibnamefont{and} \bibinfo{author}{\bibfnamefont{G.}~\bibnamefont{Zerah}},
  \bibinfo{journal}{Phys. Rev. A} \textbf{\bibinfo{volume}{46}},
  \bibinfo{pages}{5130} (\bibinfo{year}{1992}),
  \urlprefix\url{https://link.aps.org/doi/10.1103/PhysRevA.46.5130}.

\bibitem[{\citenamefont{Shukla and Eliasson}(2011)}]{RevModPhys.83.885}
\bibinfo{author}{\bibfnamefont{P.~K.} \bibnamefont{Shukla}} \bibnamefont{and}
  \bibinfo{author}{\bibfnamefont{B.}~\bibnamefont{Eliasson}},
  \bibinfo{journal}{Rev. Mod. Phys.} \textbf{\bibinfo{volume}{83}},
  \bibinfo{pages}{885} (\bibinfo{year}{2011}),
  \urlprefix\url{https://link.aps.org/doi/10.1103/RevModPhys.83.885}.

\bibitem[{\citenamefont{Palade}(2016)}]{doi:10.1063/1.4958324}
\bibinfo{author}{\bibfnamefont{D.~I.} \bibnamefont{Palade}},
  \bibinfo{journal}{Physics of Plasmas} \textbf{\bibinfo{volume}{23}},
  \bibinfo{pages}{074504} (\bibinfo{year}{2016}),
  \eprint{https://doi.org/10.1063/1.4958324},
  \urlprefix\url{https://doi.org/10.1063/1.4958324}.

\bibitem[{\citenamefont{Karasiev and Trickey}(2012)}]{KARASIEV20122519}
\bibinfo{author}{\bibfnamefont{V.}~\bibnamefont{Karasiev}} \bibnamefont{and}
  \bibinfo{author}{\bibfnamefont{S.}~\bibnamefont{Trickey}},
  \bibinfo{journal}{Computer Physics Communications}
  \textbf{\bibinfo{volume}{183}}, \bibinfo{pages}{2519 }
  (\bibinfo{year}{2012}), ISSN \bibinfo{issn}{0010-4655},
  \urlprefix\url{http://www.sciencedirect.com/science/article/pii/S0010465512002287}.

\bibitem[{\citenamefont{Benguria et~al.}(1981)\citenamefont{Benguria, Brezis,
  and Lieb}}]{Benguria1981}
\bibinfo{author}{\bibfnamefont{R.}~\bibnamefont{Benguria}},
  \bibinfo{author}{\bibfnamefont{H.}~\bibnamefont{Brezis}}, \bibnamefont{and}
  \bibinfo{author}{\bibfnamefont{E.~H.} \bibnamefont{Lieb}},
  \bibinfo{journal}{Communications in Mathematical Physics}
  \textbf{\bibinfo{volume}{79}}, \bibinfo{pages}{167} (\bibinfo{year}{1981}),
  ISSN \bibinfo{issn}{1432-0916},
  \urlprefix\url{https://doi.org/10.1007/BF01942059}.

\bibitem[{\citenamefont{Bohm}(1952)}]{PhysRev.85.166}
\bibinfo{author}{\bibfnamefont{D.}~\bibnamefont{Bohm}}, \bibinfo{journal}{Phys.
  Rev.} \textbf{\bibinfo{volume}{85}}, \bibinfo{pages}{166}
  (\bibinfo{year}{1952}),
  \urlprefix\url{https://link.aps.org/doi/10.1103/PhysRev.85.166}.

\bibitem[{\citenamefont{Kirzhnits}(1957)}]{Kirzhnits1957}
\bibinfo{author}{\bibfnamefont{D.~A.} \bibnamefont{Kirzhnits}}
  (\bibinfo{year}{1957}),
  \urlprefix\url{https://www.osti.gov/servlets/purl/4344406}.

\bibitem[{\citenamefont{Karasiev et~al.}(2013)\citenamefont{Karasiev,
  Chakraborty, Shukruto, and Trickey}}]{PhysRevB.88.161108}
\bibinfo{author}{\bibfnamefont{V.~V.} \bibnamefont{Karasiev}},
  \bibinfo{author}{\bibfnamefont{D.}~\bibnamefont{Chakraborty}},
  \bibinfo{author}{\bibfnamefont{O.~A.} \bibnamefont{Shukruto}},
  \bibnamefont{and} \bibinfo{author}{\bibfnamefont{S.~B.}
  \bibnamefont{Trickey}}, \bibinfo{journal}{Phys. Rev. B}
  \textbf{\bibinfo{volume}{88}}, \bibinfo{pages}{161108}
  (\bibinfo{year}{2013}),
  \urlprefix\url{https://link.aps.org/doi/10.1103/PhysRevB.88.161108}.

\bibitem[{\citenamefont{Yan}(2015)}]{PhysRevB.91.115416}
\bibinfo{author}{\bibfnamefont{W.}~\bibnamefont{Yan}}, \bibinfo{journal}{Phys.
  Rev. B} \textbf{\bibinfo{volume}{91}}, \bibinfo{pages}{115416}
  (\bibinfo{year}{2015}),
  \urlprefix\url{https://link.aps.org/doi/10.1103/PhysRevB.91.115416}.

\bibitem[{\citenamefont{Cirac\`{\i} and Della~Sala}(2016)}]{PhysRevB.93.205405}
\bibinfo{author}{\bibfnamefont{C.}~\bibnamefont{Cirac\`{\i}}} \bibnamefont{and}
  \bibinfo{author}{\bibfnamefont{F.}~\bibnamefont{Della~Sala}},
  \bibinfo{journal}{Phys. Rev. B} \textbf{\bibinfo{volume}{93}},
  \bibinfo{pages}{205405} (\bibinfo{year}{2016}),
  \urlprefix\url{https://link.aps.org/doi/10.1103/PhysRevB.93.205405}.

\bibitem[{\citenamefont{Akbari-Moghanjoughi}(2015)}]{doi:10.1063/1.4907167}
\bibinfo{author}{\bibfnamefont{M.}~\bibnamefont{Akbari-Moghanjoughi}},
  \bibinfo{journal}{Physics of Plasmas} \textbf{\bibinfo{volume}{22}},
  \bibinfo{pages}{022103} (\bibinfo{year}{2015}),
  \eprint{https://doi.org/10.1063/1.4907167},
  \urlprefix\url{https://doi.org/10.1063/1.4907167}.

\bibitem[{\citenamefont{Shukla and Eliasson}(2006)}]{PhysRevLett.96.245001}
\bibinfo{author}{\bibfnamefont{P.~K.} \bibnamefont{Shukla}} \bibnamefont{and}
  \bibinfo{author}{\bibfnamefont{B.}~\bibnamefont{Eliasson}},
  \bibinfo{journal}{Phys. Rev. Lett.} \textbf{\bibinfo{volume}{96}},
  \bibinfo{pages}{245001} (\bibinfo{year}{2006}),
  \urlprefix\url{https://link.aps.org/doi/10.1103/PhysRevLett.96.245001}.

\bibitem[{\citenamefont{Shaikh and Shukla}(2007)}]{PhysRevLett.99.125002}
\bibinfo{author}{\bibfnamefont{D.}~\bibnamefont{Shaikh}} \bibnamefont{and}
  \bibinfo{author}{\bibfnamefont{P.~K.} \bibnamefont{Shukla}},
  \bibinfo{journal}{Phys. Rev. Lett.} \textbf{\bibinfo{volume}{99}},
  \bibinfo{pages}{125002} (\bibinfo{year}{2007}),
  \urlprefix\url{https://link.aps.org/doi/10.1103/PhysRevLett.99.125002}.

\bibitem[{\citenamefont{Wang et~al.}(1999)\citenamefont{Wang, Govind, and
  Carter}}]{PhysRevB.60.16350}
\bibinfo{author}{\bibfnamefont{Y.~A.} \bibnamefont{Wang}},
  \bibinfo{author}{\bibfnamefont{N.}~\bibnamefont{Govind}}, \bibnamefont{and}
  \bibinfo{author}{\bibfnamefont{E.~A.} \bibnamefont{Carter}},
  \bibinfo{journal}{Phys. Rev. B} \textbf{\bibinfo{volume}{60}},
  \bibinfo{pages}{16350} (\bibinfo{year}{1999}),
  \urlprefix\url{https://link.aps.org/doi/10.1103/PhysRevB.60.16350}.

\bibitem[{\citenamefont{Huang and Carter}(2010)}]{PhysRevB.81.045206}
\bibinfo{author}{\bibfnamefont{C.}~\bibnamefont{Huang}} \bibnamefont{and}
  \bibinfo{author}{\bibfnamefont{E.~A.} \bibnamefont{Carter}},
  \bibinfo{journal}{Phys. Rev. B} \textbf{\bibinfo{volume}{81}},
  \bibinfo{pages}{045206} (\bibinfo{year}{2010}),
  \urlprefix\url{https://link.aps.org/doi/10.1103/PhysRevB.81.045206}.

\bibitem[{\citenamefont{Moldabekov et~al.}()\citenamefont{Moldabekov, Bonitz,
  and Ramazanov}}]{doi:10.1002/ctpp.201700113}
\bibinfo{author}{\bibfnamefont{Z.}~\bibnamefont{Moldabekov}},
  \bibinfo{author}{\bibfnamefont{M.}~\bibnamefont{Bonitz}}, \bibnamefont{and}
  \bibinfo{author}{\bibfnamefont{T.}~\bibnamefont{Ramazanov}},
  \bibinfo{journal}{Contributions to Plasma Physics}
  \textbf{\bibinfo{volume}{57}}, \bibinfo{pages}{499} (????),
  \eprint{https://onlinelibrary.wiley.com/doi/pdf/10.1002/ctpp.201700113},
  \urlprefix\url{https://onlinelibrary.wiley.com/doi/abs/10.1002/ctpp.201700113}.

\bibitem[{\citenamefont{Hohenberg and Kohn}(1964)}]{PhysRev.136.B864}
\bibinfo{author}{\bibfnamefont{P.}~\bibnamefont{Hohenberg}} \bibnamefont{and}
  \bibinfo{author}{\bibfnamefont{W.}~\bibnamefont{Kohn}},
  \bibinfo{journal}{Phys. Rev.} \textbf{\bibinfo{volume}{136}},
  \bibinfo{pages}{B864} (\bibinfo{year}{1964}),
  \urlprefix\url{https://link.aps.org/doi/10.1103/PhysRev.136.B864}.

\bibitem[{\citenamefont{Bonitz}(2015)}]{9783319241210}
\bibinfo{author}{\bibfnamefont{M.}~\bibnamefont{Bonitz}},
  \emph{\bibinfo{title}{Quantum Kinetic Theory}}
  (\bibinfo{publisher}{Springer}, \bibinfo{year}{2015}),
  \urlprefix\url{https://www.amazon.com/Quantum-Kinetic-Theory-Michael-Bonitz-ebook/dp/B018AENOS2?SubscriptionId=AKIAIOBINVZYXZQZ2U3A&tag=chimbori05-20&linkCode=xm2&camp=2025&creative=165953&creativeASIN=B018AENOS2}.

\bibitem[{\citenamefont{Bohm and Vigier}(1954)}]{PhysRev.96.208}
\bibinfo{author}{\bibfnamefont{D.}~\bibnamefont{Bohm}} \bibnamefont{and}
  \bibinfo{author}{\bibfnamefont{J.~P.} \bibnamefont{Vigier}},
  \bibinfo{journal}{Phys. Rev.} \textbf{\bibinfo{volume}{96}},
  \bibinfo{pages}{208} (\bibinfo{year}{1954}),
  \urlprefix\url{https://link.aps.org/doi/10.1103/PhysRev.96.208}.

\bibitem[{\citenamefont{Giuliani and Vignale}(2008)}]{9780521527965}
\bibinfo{author}{\bibfnamefont{G.}~\bibnamefont{Giuliani}} \bibnamefont{and}
  \bibinfo{author}{\bibfnamefont{G.}~\bibnamefont{Vignale}},
  \emph{\bibinfo{title}{Quantum Theory of the Electron Liquid}}
  (\bibinfo{publisher}{Cambridge University Press}, \bibinfo{year}{2008}), ISBN
  \bibinfo{isbn}{0521527961},
  \urlprefix\url{https://www.amazon.com/Quantum-Theory-Electron-Gabriele-Giuliani/dp/0521527961?SubscriptionId=AKIAIOBINVZYXZQZ2U3A&tag=chimbori05-20&linkCode=xm2&camp=2025&creative=165953&creativeASIN=0521527961}.

\bibitem[{\citenamefont{Gottlieb and Orszag}(1987)}]{9780898710236}
\bibinfo{author}{\bibfnamefont{D.}~\bibnamefont{Gottlieb}} \bibnamefont{and}
  \bibinfo{author}{\bibfnamefont{S.~A.} \bibnamefont{Orszag}},
  \emph{\bibinfo{title}{Numerical Analysis of Spectral Methods : Theory and
  Applications (CBMS-NSF Regional Conference Series in Applied Mathematics)}}
  (\bibinfo{publisher}{Society for Industrial and Applied Mathematics},
  \bibinfo{year}{1987}), ISBN \bibinfo{isbn}{0898710235},
  \urlprefix\url{https://www.amazon.com/Numerical-Analysis-Spectral-Methods-Applications/dp/0898710235?SubscriptionId=AKIAIOBINVZYXZQZ2U3A&tag=chimbori05-20&linkCode=xm2&camp=2025&creative=165953&creativeASIN=0898710235}.

\bibitem[{\citenamefont{Levy}(1979)}]{Levy6062}
\bibinfo{author}{\bibfnamefont{M.}~\bibnamefont{Levy}},
  \bibinfo{journal}{Proceedings of the National Academy of Sciences}
  \textbf{\bibinfo{volume}{76}}, \bibinfo{pages}{6062} (\bibinfo{year}{1979}),
  ISSN \bibinfo{issn}{0027-8424},
  \eprint{http://www.pnas.org/content/76/12/6062.full.pdf},
  \urlprefix\url{http://www.pnas.org/content/76/12/6062}.

\bibitem[{\citenamefont{Brack}(1993)}]{RevModPhys.65.677}
\bibinfo{author}{\bibfnamefont{M.}~\bibnamefont{Brack}}, \bibinfo{journal}{Rev.
  Mod. Phys.} \textbf{\bibinfo{volume}{65}}, \bibinfo{pages}{677}
  (\bibinfo{year}{1993}),
  \urlprefix\url{https://link.aps.org/doi/10.1103/RevModPhys.65.677}.

\end{thebibliography}

\end{document}